\documentclass[aps,preprint,preprintnumbers,nofootinbib,showpacs]{revtex4}

\usepackage{amsmath}
\usepackage{amssymb}
\usepackage{dsfont}
\usepackage{color}
\usepackage{graphicx,accents}
\usepackage[normalem]{ulem}

\def\bea{\begin{eqnarray}}
\def\eea{\end{eqnarray}}
\def\sea{\nonumber \\&&}
\def\lla{\left\langle}
\def\rra{\right\rangle}
\def\za{\alpha}
\def\zc{\gamma}
\def\zb{\beta}

\newcommand{\shi}{_{\!\ssc (\!\chi\!)}} 
\newcommand{\shia}{_{\!\ssc (\!\chi\!)A}}
\newcommand{\shib}{_{\!\ssc (\!\chi\!)B}}
\newcommand{\vsi}{_{\!\ssc (\!\varsigma\!)}} 
\newcommand{\vsia}{_{\!\ssc (\!\varsigma\!)A}}

\def\nfac{n_{\!\ssc 1}!\,n_{\!\ssc 2}!\,n_{\!\ssc 3}!\,n_{\!\ssc 4}!}

\def\ssc{\scriptscriptstyle}
\def\lsim{\mathrel{\raise.3ex\hbox{$<$\kern-.75em\lower1ex\hbox{$\sim$}}} }
\def\gsim{\mathrel{\raise.3ex\hbox{$>$\kern-.75em\lower1ex\hbox{$\sim$}}} }
\makeatletter
\DeclareRobustCommand{\cev}[1]{%
  \mathpalette\do@cev{#1}%
}
\newcommand{\do@cev}[2]{%
  \fix@cev{#1}{+}%
  \reflectbox{$\m@th#1\vec{\reflectbox{$\fix@cev{#1}{-}\m@th#1#2\fix@cev{#1}{+}$}}$}%
  \fix@cev{#1}{-}%
}
\newcommand{\fix@cev}[2]{%
  \ifx#1\displaystyle
    \mkern#23mu
  \else
    \ifx#1\textstyle
      \mkern#23mu
    \else
      \ifx#1\scriptstyle
        \mkern#22mu
      \else
        \mkern#22mu
      \fi
    \fi
  \fi
}

\makeatother
\begin{document}
\preprint{{\vbox{\hbox{NCU-HEP-k084}
\hbox{Jan 2020}
}}}
\vspace*{.7in}


\title{Covariant Quantum Mechanics and Quantum Spacetime
\vspace*{.2in}}

\author{Suzana Bedi\'c,} 

\address{ICRANet, P.le della Repubblica 10, 65100 Pescara, Italy,\\ and
ICRA and University of Rome ``Sapienza'', Physics Department, P.le
A. Moro 5, 00185 Rome, Italy
\vspace*{.3in} }
\author{Otto C. W. Kong, and Hock King Ting}

\address{ Department of Physics and Center for High Energy and High Field Physics,
National Central University, Chung-li, Taiwan 32054  \\
}


\begin{abstract}
\vspace*{.3in}
We present in the article the formulation of a version
of Lorentz covariant quantum mechanics based on a group
theoretical construction from a Heisenberg-Weyl symmetry
with position and momentum operators transforming as
Minkowski four-vectors under the Lorentz symmetry. The
basic representation is identified as a coherent state
representation, essentially an irreducible component 
of the regular representation, with the matching 
representation of an extension of the group $C^*$-algebra 
giving the algebra of observables. The key feature of the 
formulation is that it is not unitary but pseudo-unitary, 
exactly in the same sense as the Minkowski spacetime  
representation. Explicit wavefunction description is
given without any restriction of the variable domains,
yet with a finite integral inner product. The associated 
covariant harmonic oscillator Fock state basis has all 
the standard properties in exact analog to those of a 
harmonic oscillator with Euclidean position and momentum 
operators of any `dimension'. Galilean limit of the 
Lorentz symmetry and the classical limit of the Lorentz 
covariant framework are retrieved rigorously through  
appropriate symmetry contractions of  the algebra and 
its representation, including the dynamics described 
through the symmetry of the phase space, given both 
in terms of real/complex number coordinates and 
noncommutative operator coordinates. The latter
gives an explicit picture of the (projective)
Hilbert space as a quantum/noncommutative spacetime.

\end{abstract}


\maketitle

\section{Introduction}
The formulation of a fully Lorentz covariant version of quantum 
mechanics with position and momentum operators $\hat{X}_\mu$
and $\hat{P}_\mu$ transforming as Minkowski four-vectors
has been around since the early days of quantum mechanics.
A naive thinking would be to take the representation of those
operators as $x_\mu$ and $-i\hbar \partial_{x^\mu}$, acting on 
the wavefunctions $\psi(x^\mu)$ with the simple inner product
giving the squared integral norm and a unitary Schr\"odinger 
evolution under the Einstein proper time $\tau$. Explicit
group theoretical picture of that, under what we called
$H_{\!\ssc R}(1,3)$ symmetry, has been available since the
sixties \cite{Z,J}.  There are numerous studies on such 
a theory and its variants which we refrained from quoting.
The results are not quite satisfactory. Here, we revisit the
subject matter with very different perspectives, and give
a formulation based on a pseudo-unitary representation
which, in our opinion, beautifully resolves all the problems.

The difficulties of the usual unitary approach are particularly
well illustrated in the analysis of the covariant harmonic 
oscillator system. The importance of the harmonic oscillator
problem in any setting can hardly be overstated.  In the 
standard `non-relativistic' quantum mechanics, the Fock states 
are one of the most useful orthonormal basis for the Hilbert 
space and the latter, as the space of rapidly decreasing 
functions spanned by their wavefunctions, rigorously gives
the states on which the position and momentum operators
can be truly Hermitian \cite{M}. However, solutions to the 
covariant harmonic oscillator problem, under a unitary or 
non-unitary formulation, have difficulties with less than
nice expected Lorentz transformation properties or
wavefunctions with divergence issues \cite{Z,B}. We just
revisit the problem with the idea of taking a pseudo-unitary 
representation, which is like a natural extension of the
Minkowski spacetime, as a representation of the Lorentz
symmetry, seeing the lack of full unitarity as a basic
signature of spacetime physics \cite{082}. Our solutions,
given in terms of  $\hat{X}_a$ and $\hat{P}_a$ as
$x_a$ and $-i\hbar \partial_{x^a}$, $a=1,2,3,4$, and
the anti-Hermitian $\hat{X}_{\!\ssc 0} =i \hat{X}_{\!\ssc 4}$ 
and $\hat{P}_{\!\ssc 0} =i \hat{P}_{\!\ssc 4}$, have been
established to have the desirable Lorentz transformation
properties while being free from any divergence issue \cite{083}.
The usual inner product is not the right, Lorentz invariant,
one though. The right one introduced is pseudo-unitary,
giving norms that can be spacelike (+ve), timelike (-ve),
or lightlike (0).  

With the lesson learned from our solutions to the covariant
harmonic oscillator problem, here we look at the exact
formulation of the covariant quantum mechanics itself
from a coherent state point of view. The current study
targets two important aspects, and achieves the results
we want.

Firstly, our  group had implemented a quantum relativity
symmetry group theoretical perspective to formulate the 
full dynamical theory of the familiar quantum mechanics 
with rigorous classical limit given as a Newtonian theory,
obtained through a contraction of the relativity symmetry 
applied to the specific representation. The latter is taken
as essentially an irreducible component of the regular 
representation of $H(3)$, the Heisenberg-Weyl group. The 
full quantum relativity symmetry, denoted $\tilde{G}(3)$, 
can naturally be seen as a $U(1)$ central extension of the 
Galilean symmetry. $H_{\!\ssc R}(3)$ is (or is isomorphic to) 
its subgroup, left after the `time-translation' is taken out. 
A $H(3)$ representation is a spin zero, time independent, 
representation of $\tilde{G}(3)$. The representation is 
really the one of the canonical coherent states. The 
matching representation of the group $C^*$-algebra, further 
extended to a proper class of distributions, gives the
observable algebra as functions, and distributions, of 
$\hat{X}_i$ and $\hat{P}_i$, essentially as given by 
the Weyl-Wigner-Groenewold-Moyal(WWGM) formulation. The 
operators $\za(p_i\star,x_i\star)= \za(p_i,x_i)\star$ 
act as differential operators on the coherent state 
wavefunctions $\phi(p^i,x^i)$ by the Moyal star-product 
$\za\star\phi$; $\za\star\zb\star = (\za\star\zb)\star$.  
A detailed study emphasizing on a quantum space model 
is given in Ref. \cite{066}, and on the noncommutative 
geometric perspectives in Ref.\cite{070}. We want to 
give the analogous formulation for $H_{\!\ssc R}(1,3)$ 
as an upper level quantum relativity symmetry along 
the contraction chain \cite{030,071}, and hence give 
also the first quantum/noncommutative spacetime model 
solidly based on the known physics.

Secondly, we want to look at the covariant theory of 
Ref.\cite{083} in relation to the group theoretical
formulation from $H_{\!\ssc R}(1,3)$ group symmetry to 
understand a more physical picture of it in terms of 
the Minkowski four-vector variables $x^\mu$ and $p^\mu$ 
and operators $x_\mu\star$ and $p_\mu\star$, hence 
circumventing the difficulties of a description based 
on wavefunctions of the type $\phi(x^a)$ or $\phi(p^a,x^a)$. 
  
The current paper reports the success of the work we set out 
to do. To follow the formulation presented in Ref.\cite{083}, 
we start with the unitary representation of $H_{\!\ssc R}(4)$,
illustrating the irreducible component of the regular 
representation of $H(4)$, in Sec.\ref{sec2}. quite some 
details are given, with somewhat complicated-looking notation, 
since the parallel details for  $H(3)$ had not been explicitly 
shown in Ref.\cite{070}, which have direct analogs in the case 
of $H(1,3)$ and are needed to understand the contraction of 
the Lorentz symmetry to the Galilean one. The explicit 
$H_{\!\ssc R}(1,3)$ picture  is presented in Sec.\ref{sec3}, 
in which we go all the way to present also the coherent states 
for the $H_{\!\ssc R}(1,3)$ in the same abstract Hilbert space,
showing the representation as essentially the one corresponding 
to the regular representation of $H(1,3)$, which is hence still
pseudo-unitary with the Lorentz invariant physical inner product. 
Without going through the path via $H_{\!\ssc R}(4)$, it is 
difficult to see that important pseudo-unitarity and its 
consequence. In particular, we get the wavefunctions
$\tilde\phi(p^\mu,x^\mu)$ with the finite integral inner 
product, all having nice enough analytical properties otherwise 
difficult to obtain. Sec.\ref{sec4} deals with the Lorentz 
to Galilean contraction of the representation. Sec.\ref{sec5}
is devoted to the WWGM framework or the observable algebra, 
focusing on the symmetry transformations and the dynamics as 
a specific case such a symmetry flow with the real parameter 
characterizing transformation corresponding to an evolution 
parameter which is taken as the proper time in the case. 
Sec.\ref{sec6} gives the direct contraction at the Lorentz 
covariant level to a classical limit. Discussions and 
conclusions are given in the last section. In the appendix 
we summarize the results obtained in the Fock state basis,
some of which are used in Sec.\ref{sec3}.

\section{The Representation from Irreducible Representations of
\boldmath $H_{\!\ssc R}(4)$ \label{sec2}}
We give the Lie algebra for $H_{\!\ssc R}(1,3)$ as 
\bea &&
[J_{\mu\nu}, J_{\rho\sigma}] 
= 2i \left( \eta_{\nu\sigma} J_{\mu\rho} 
  + \eta_{\mu\rho} J_{\nu\sigma} - \eta_{\mu\sigma} J_{\nu\rho} 
   -\eta_{\nu\rho} J_{\mu\sigma}\right) \;,
\sea
[J_{\mu\nu}, Y_\rho] = 2 i  \left( \eta_{\mu\rho} Y_{\nu} 
   - \eta_{\nu\rho} Y
   _{\mu} 
  \right) \;,
  \sea
[J_{\mu\nu}, E_\rho] = 2 i \left( \eta_{\mu\rho} E_{\nu} 
   - \eta_{\nu\rho} E_{\mu} 
  \right) \;,
\sea
[Y_\mu, E_\nu] =2i \eta_{\mu\nu} I\;,
\label{h13}
\eea
where $\eta_{\mu\nu}=\mbox{diag}\{-1,1,1,1\}$. The choice of 
notation with $Y_\mu$ corresponding essentially to spacetime 
position observables and $E_\mu$ to energy-momentum  observables 
is somewhat unusual. The reason for it should be clear from the 
analysis below.  Notice that the generators are all taken to have
no physical dimension, and the factor $2$ corresponds to $\hbar$ 
in the chosen units, which is at least convenient for the coherent 
state formulation \cite{070}. In terms of the group element
$g(p^\mu, x^\mu, \theta,\Lambda^{\!\mu}_\nu)$, 
we have (with the indices suppressed)
\bea &&
g(p', x', \theta', \Lambda')  g(p, x, \theta,\Lambda)
\sea\hspace*{.2in}
= g\left(p'+ \Lambda' p, x'+\Lambda' x, 
     \theta'\!+\theta\!- x' \Lambda' p + p'\Lambda' x, 
     \Lambda'\Lambda \right) .
\label{group}
\eea
The story is an extension of what has been done in Ref.\cite{066,070} 
for $H_{\!\ssc R}(3)=H\!(3) \rtimes SO(3)$ to the framework of
\bea
H_{\!\ssc R}(1,3)=H\!(1,3) \rtimes SO(1,3) \;,
\eea
the focus of which, for the spin zero case here, is only on 
the irreducible representation of the Heisenberg-Weyl 
symmetry $H\!(1,3)$ and $H\!(3)$. A key point of difference 
between the two cases is that $SO(1,3)$ is noncompact, the 
finite dimensional representations of which, as direct 
extension of those compact ones of $SO(3)$, are pseudo-unitary 
instead of unitary. The basis of that pseudo-unitarity is 
the indefinite Minkowski norm associated with the metric
$\eta_{\mu\nu}$ extending the Euclidean $\delta_{ij}$
 \cite{082,083}. In the case of $H_{\!\ssc R}(3)$, the 
representation is naturally an irreducible component of the
regular representation of $H\!(3)$, which all can be seen
actually as physically equivalent. The regular representation 
for a Lie group is however unitary, at least in the sense that
the generators of the Lie algebra represented as invariant
vector fields are naively Hermitian, {\em i.e.} with respect 
to the usual inner product on the functional space. Inspired 
by the analysis of the corresponding Lorentz covariant
harmonic oscillator problem \cite{083}, we take here a 
simple approach used there to construct the pseudo-unitary 
representation,  however, in a coherent state basis with 
the wavefunctions naturally serving as the description 
of the states under the  WWGM formalism, as has been
 done for the case of $H\!(3)$ group \cite{070}.
 
Our basic approach to the proper pseudo-unitary representation
is essentially that of the `Weyl trick' \cite{G}, using the 
relation between the representations of $SO(1,3)$ and that of 
$SO(4)$ sharing the same complexification. We first present 
the results from a harmonic analysis of Heisenberg-Weyl groups 
adopted to our case of $H(4)$ \cite{T}. The left regular 
representation is written, in $\hbar=2$ units, as 
$U (p^a,x^a,\theta)
   =e^{i(p^a Y^{\!\ssc L}_a -x^a E^{\!\ssc L}_a+\theta I^{\!\ssc L})}$,
where 
\bea
Y^{\!\ssc L}_a &=&i  x_a  \partial_\theta +  i \partial_{p^a}\;,
\nonumber \\
E^{\!\ssc L}_a &=& i  p_a \partial_\theta -  i \partial_{x^a}\;,
\nonumber \\
I^{\!\ssc L} &=& i \partial_\theta \;,
\eea
are the left-invariant vector fields. In a unitary irreducible
representation, all of which are contained in the regular 
representation, the central generator $I$ has  to be represented 
by a real multiple of identity. We write the one parameter series 
$U_{\!\ssc\varsigma}$ ($\varsigma\ne 0$) of representations for 
the generators as Hermitian operators as
$\{\hat{Y}_{\!\ssc \varsigma}^{\!\ssc L}, \hat{E}_{\!\ssc \varsigma}^{\!\ssc L}, \varsigma \hat{I}\}$, 
where $\hat{I}$ is the identity operator and 
$[\hat{Y}^{\!\ssc L}_{\!\ssc \varsigma a}, \hat{E}^{\!\ssc L}_{\!\ssc \varsigma b}]
   =2 i\varsigma\delta_{ab} \hat{I}$.
The ${U}^{\!\ssc L}_{\!\ssc\varsigma}$ set can be considered the set of 
equivalence classes of irreducible unitary representations with nonzero 
Plancherel measure. The limit of $U^{\!\ssc L}_{\!\ssc \varsigma}$ as 
$\varsigma \to 0$ gives the whole set of irreducible one dimensional 
representations. The latter set has zero Plancherel measure and
together with the $U^{\!\ssc L}_{\!\ssc \varsigma}$ exhausts all
equivalence classes of irreducible representations. Based on the 
measure, one should consider the expansion
\bea \label{za}
\za(p^a, x^a,\theta)=  \frac{1}{(2\pi)^{\frac{1}{2}}}\int \!\! d\varsigma \, 
   \za_{\ssc\varsigma} (p^a, x^a) \,e^{-i\varsigma\theta} |\varsigma|^n\;,
\eea
$n=1+3$ here, given as the inverse Fourier-Plancherel  transform. 
The actions of the left-invariant vector fields on $\za(p,x,\theta)$ 
in the form of Eq.(\ref{za}) are given by their actions on 
$\za_{\ssc \varsigma}(p,x)e^{-i\varsigma\theta}$ parts as 
${\varsigma}x+i \partial_{p}$, ${\varsigma}p-i \partial_{x}$, 
and $\varsigma$, respectively.  Here, and below, we suppress 
the indices wherever it is unambiguous. We can see that the 
action at each $\varsigma \ne 0$ corresponds exactly to the  
${U}^{\!\ssc L}_{\!\ssc\varsigma}$ representation 
with the generators represented by  $\{ \hat{Y}^{\!\ssc L}_{\!\ssc \varsigma}, 
  \hat{E}^{\!\ssc L}_{\!\ssc \varsigma},  \varsigma \hat{I}\}$. 
That is the reduction of the regular representation into irreducible
components. For positive values of  $\varsigma$, one can introduce 
the $\varsigma$ independent operators
\bea && \label{genop}
\hat{X}^{\!\ssc L}_{\!\ssc (\!\varsigma\!)} \equiv \frac{1}{\sqrt{\varsigma}} \hat{Y}^{\!\ssc L}_{\ssc \!\varsigma}
  = x_{\ssc (\!\varsigma\!)}  +  i \partial_{p_{\ssc (\!\varsigma\!)}}\;,
\sea
\hat{P}^{\!\ssc L}_{\!\ssc (\!\varsigma\!)} \equiv \frac{1}{\sqrt{\varsigma}} \hat{E}^{\!\ssc L}_{\ssc \!\varsigma}
  = p_{\ssc (\!\varsigma\!)}  -  i \partial_{x_{\ssc (\!\varsigma\!)}}\;,
\eea
where we have $x_{\ssc (\!\varsigma\!)}= \sqrt{\varsigma} x$ 
and $p_{\ssc (\!\varsigma\!)}= \sqrt{\varsigma} p$. 
${U}^{\!\ssc L}_{\!\ssc\varsigma}(p,x,\theta)$ is then given by
$e^{i(p_{\ssc (\!\varsigma\!)}\hat{X}^{\!\ssc L}_{\!\ssc (\!\varsigma\!)}
   -x_{\ssc (\!\varsigma\!)}\hat{P}^{\!\ssc L}_{\!\ssc (\!\varsigma\!)}
    +\theta_{\ssc (\!\varsigma\!)} \hat{I})}$,
with $\theta_{\ssc (\!\varsigma\!)}={\varsigma} {\theta}$, 
hence in a form formally independent of $\varsigma$. 
$\hat{X}^{\!\ssc L}_{\!\ssc (\!\varsigma\!)}$ and
 $\hat{P}^{\!\ssc L}_{\!\ssc (\!\varsigma\!)}$ are still $SO(4)$ 
vectors, and so are $p_{\ssc (\!\varsigma\!)}$ and 
$x_{\ssc (\!\varsigma\!)}$. The $(\!\varsigma\!)$ index becomes 
completely dummy and analysis based on the new operators and 
the parameters would be independent of $\varsigma$ so long 
as we are looking only at a particular irreducible representation.  
One can even simply drop it. From a physics perspective, we 
have absorbed the value of $\varsigma$ by a choice of 
physical unit for measuring the observables corresponding 
to $Y$ and $E$, here all in unit of  $\sqrt{\varsigma}$. 
 For $\varsigma$ being negative\footnote{
From the physical point of view, the representations corresponding 
to different value of $\varsigma$ can be seen as describing the 
same physics.  The parameter $\varsigma$ may then be taken as the 
physical constant $\frac{\hbar c^2}{2}$.  And for that matter, 
$\varsigma$ cannot be negative. Physicists identify the symmetry 
algebra from a relevant representation with 
$\hat{X}^{\!\ssc L}_{\!\ssc (\!\varsigma\!)}$ and 
$\hat{P}^{\!\ssc L}_{\!\ssc (\!\varsigma\!)}$ 
as the position and momentum observables satisfying
$[\hat{X}^{\!\ssc L}_{\!\ssc (\!\varsigma\!)a},\hat{P}^{\!\ssc L}_{\!\ssc (\!\varsigma\!)b}]
=2i\delta_{ab}$,  in the $\hbar=2$ units. However, 
the mathematical the case of a product of two 
representations with different $\varsigma$ values 
may have interesting physics implications if composite 
physical system corresponding to that exists in nature.  
},
 we should switch $\hat{Y}^{\!\ssc L}_{\!\ssc \varsigma}$ 
with $\hat{E}^{\!\ssc L}_{\!\ssc \varsigma}$ first;   {\em i.e.} we take  
\bea &&
\hat{X}^{\!\ssc L}_{\!\ssc (\!\varsigma\!)} 
           \equiv \frac{1}{\sqrt{|\varsigma|}} \hat{E}^{\!\ssc L}_{\ssc \!\varsigma}
  = x_{\ssc (\!\varsigma\!)}  +  i \partial_{p_{\ssc (\!\varsigma\!)}}\;,
\sea\nonumber
\hat{P}^{\!\ssc L}_{\!\ssc (\!\varsigma\!)} 
              \equiv \frac{1}{\sqrt{|\varsigma|}} \hat{Y}^{\!\ssc L}_{\ssc \!\varsigma}
  = p_{\ssc (\!\varsigma\!)}  -  i \partial_{x_{\ssc (\!\varsigma\!)}}\;,
\eea
achieved by taking $x_{\ssc (\!\varsigma\!)}= -\sqrt{|\varsigma|} p$ 
and $p_{\ssc (\!\varsigma\!)}= -\sqrt{|\varsigma|} x$.  
The result still maintains
$U^{\!\ssc L}_{\varsigma}(p_{\ssc (\!\varsigma\!)},   x_{\ssc (\!\varsigma\!)},
  \theta_{\ssc (\!\varsigma\!)})
    = e^{i(p_{\ssc (\!\varsigma\!)}\hat{X}^{\!\ssc L}_{\!\ssc (\!\varsigma\!)}
   -x_{\ssc (\!\varsigma\!)}\hat{P}^{\!\ssc L}_{\!\ssc (\!\varsigma\!)}
    +\theta_{\ssc (\!\varsigma\!)} \hat{I})}$ with
$[\hat{X}^{\!\ssc L}_{\!\ssc (\!\varsigma\!)a},\hat{P}^{\!\ssc L}_{\!\ssc (\!\varsigma\!)b}]
 = 2i\delta_{ab} \hat{I}$.
$\varsigma$ can actually be seen as the eigenvalue of $I$, 
essentially the Casimir operator. The semidirect product structure 
of the $H_{\!\ssc R}(4)=H\!(4) \rtimes SO(4)$ says that with each
irreducible unitary representation of the subgroup 
$H\!(4) \rtimes S_{\hat{O}}$, where $S_{\hat{O}} \subseteq SO(4)$ 
is the stability subgroup for an orbit $\hat{O}$ of $SO(4)$ in the 
space of equivalent classes of irreducible unitary representations 
of $H(4)$, one can associate an induced representation which is
irreducible \cite{BR}.  We have seen that, apart from the set of 
measure zero, each of which only gives one-dimensional 
representations, the irreducible unitary representations are 
characterized by the nonzero value of $\varsigma$ and the
representations (though mathematically inequivalent) 
can be casted in the same form as
$U^{\!\ssc L}_{\varsigma}(p_{\ssc (\!\varsigma\!)},   x_{\ssc (\!\varsigma\!)},
  \theta_{\ssc (\!\varsigma\!)})$.
It is obvious that the representation is invariant under the $SO(4)$
transformations, hence each is an independent orbit. That is to say 
$S_{\hat{O}} = SO(4)$. The fact is of paramount importance for
unambiguously identifying the nature of the coherent states below. 
In view of the discussion above, we can see that for any of the 
$U^{\!\ssc L}_{\varsigma}(p_{\!\ssc (\!\varsigma\!)},   x_{\ssc (\!\varsigma\!)},
  \theta_{\!\ssc (\!\varsigma\!)})$
representation, we can simply write it in the simple notation
$U^{\ssc\! L}\!(p,x,\theta)$, like taking the $\varsigma=1$ 
case as a representative. That is essentially what has been 
done in Ref.\cite{070} for the $H\!(3)$ or $H_{\!\ssc R}(3)$ 
case. However, for the reason to be clear below, 
we keep the explicit $\varsigma$-notation here.

Consider the $H(4)$ group product as
\bea\label{ww}
{W}\!(p'^a\vsi, x'^a\vsi, \theta'\vsi)  {W}\!(p^a\vsi, x^a\vsi, \theta\vsi)
= {W}\!\!\left(p'^a\vsi + p^a\vsi, x'^a\vsi + x^a\vsi, 
     \theta'\vsi+\theta\vsi- \!(x'_{{\ssc (\!\varsigma\!)}_a}  p^a\vsi \!- p'_{{\ssc (\!\varsigma\!)}_a} x^a\vsi) \right) .
\eea
Following the basic approach \cite{066,070}, we introduce 
the canonical coherent states defined as
\bea
\left|p^a_{\ssc (\!\varsigma\!)},x^a_{\ssc (\!\varsigma\!)} \rra 
 \equiv {U}_{\ssc \!\varsigma}(p^a_{\!\ssc (\!\varsigma\!)},x^a_{\ssc (\!\varsigma\!)}) 
  \left|0\rra \equiv e^{-i\theta_{\ssc (\!\varsigma\!)}} 
 {U}_{\ssc \!\varsigma}(p^a_{\!\ssc (\!\varsigma\!)},x^a_{\ssc (\!\varsigma\!)},
     \theta_{\!\ssc (\!\varsigma\!)}) \left|0_{\ssc (\!\varsigma\!)}\rra \;, 
\eea
where
\bea
U_{\ssc \!\varsigma}(p^a_{\!\ssc (\!\varsigma\!)},x^a_{\ssc (\!\varsigma\!)},\theta_{\ssc (\!\varsigma\!)})
 \equiv e^{i (p^a_{\!\ssc (\!\varsigma\!)}\hat{X}_{\ssc (\!\varsigma\!)a}
  - x^a_{\ssc (\!\varsigma\!)}\hat{P}_{\ssc (\!\varsigma\!)a} 
     +\theta_{\!\ssc (\!\varsigma\!)}\hat{I})} \;
\eea
is the representation of the group element coordinated by 
$(p^a_{\!\ssc (\!\varsigma\!)},x^a_{\ssc (\!\varsigma\!)},\theta_{\ssc (\!\varsigma\!)})$ 
on the group manifold, and  $\hat{X}_{\ssc (\!\varsigma\!)}$, and 
$\hat{P}_{\ssc (\!\varsigma\!)}$ are Hermitian operators  on the 
abstract Hilbert space  $\mathcal{H}_{\ssc \varsigma}$ spanned by the
$\left|p^a_{\ssc (\!\varsigma\!)},x^a_{\ssc (\!\varsigma\!)} \rra$  
vectors; $\left|0\rra \equiv \left|0,0\rra$ being a fiducial normalized vector. 
$(p^a_{\!\ssc (\!\varsigma\!)},x^a_{\ssc (\!\varsigma\!)},\theta_{\ssc (\!\varsigma\!)})$ 
is a general element of  $H\!(4)$, and it can 
be identified with a point in the coset space of $H_{\!\ssc R}(4)/SO(4)$
 \cite{066,071}. The cyclic vector $\left|0_{\ssc (\!\varsigma\!)}\rra$ 
corresponds to the points $(0,0,\theta_{\!\ssc (\!\varsigma\!)})$
in the coset space, each of which is fixed under $SO(4)$ transformations. 
Assuming the state $\left|0_{\ssc (\!\varsigma\!)}\rra$ has zero 
expectation values of the 
$\hat{X}_{\ssc (\!\varsigma\!)}$ and $\hat{P}_{\ssc (\!\varsigma\!)}$  
operators, we get 
\bea &&
\lla p^a_{\!\ssc (\!\varsigma\!)},x^a_{\ssc (\!\varsigma\!)} 
  \left| \hat{X}_{\ssc (\!\varsigma\!)b} \right| 
    p^a_{\!\ssc (\!\varsigma\!)},x^a_{\ssc (\!\varsigma\!)} \rra 
   = 2 x_{\ssc (\!\varsigma\!)b} \;,
\sea
 \lla p^a_{\!\ssc (\!\varsigma\!)},x^a_{\ssc (\!\varsigma\!)} 
  \left| \hat{P}_{\ssc (\!\varsigma\!)b} \right| 
    p^a_{\!\ssc (\!\varsigma\!)},x^a_{\ssc (\!\varsigma\!)} \rra 
   = 2 p_{\!\ssc (\!\varsigma\!)b}  \;. 
\eea
We have the wavefunctions on the coherent state
manifold $\phi (p_{\!\ssc (\!\varsigma\!)},x_{\ssc (\!\varsigma\!)}) 
  \equiv \lla p_{\!\ssc (\!\varsigma\!)},x_{\ssc (\!\varsigma\!)} | \phi \rra$ 
(indices suppressed) with 
\bea 
 \lla p_{\!\ssc (\!\varsigma\!)},x_{\ssc (\!\varsigma\!)}  \left| \hat{X}_{\!\ssc (\!\varsigma\!)} \right|\phi \rra &=& \hat{X}^{\!\ssc L}_{\!\ssc (\!\varsigma\!)} \phi (p_{\!\ssc (\!\varsigma\!)},x_{\ssc (\!\varsigma\!)})  \;,
\nonumber \\
\lla p_{\!\ssc (\!\varsigma\!)},x_{\ssc (\!\varsigma\!)} \left| \hat{P}_{\!\ssc (\!\varsigma\!)} \right|\phi \rra &=& \hat{P}^{\!\ssc L}_{\!\ssc (\!\varsigma\!)}  \phi (p_{\!\ssc (\!\varsigma\!)},x_{\ssc (\!\varsigma\!)})    \;,
\eea
where 
\bea
\hat{X}^{\!\ssc L}_{\!\ssc (\!\varsigma\!)} &=& 
   x_{\ssc (\!\varsigma\!)}  +  i \partial_{\!p_{\ssc (\!\varsigma\!)} }\;,
\nonumber \\
\hat{P}^{\!\ssc L}_{\!\ssc (\!\varsigma\!)} &=& 
   p_{\ssc (\!\varsigma\!)}  - i \partial_{\!x_{\ssc (\!\varsigma\!)} }\;,
\label{L}
\eea
for the unitary representation on the 
$\phi (p_{\!\ssc (\!\varsigma\!)},x_{\ssc (\!\varsigma\!)})$ 
functional space satisfying
\bea \label{u-shift} &&
{U}^{\ssc\! L}_{\ssc \!\varsigma}\!(p_{\!\ssc (\!\varsigma\!)},x_{\ssc (\!\varsigma\!)}) 
  \phi(p'_{\!\ssc (\!\varsigma\!)},x'_{\ssc (\!\varsigma\!)})   \equiv
 \lla p'_{\!\ssc (\!\varsigma\!)},x'_{\ssc (\!\varsigma\!)} 
 \left|{U}_{\ssc \!\varsigma}(p_{\!\ssc (\!\varsigma\!)},x_{\ssc (\!\varsigma\!)})  \right|\phi \rra 
\sea\hspace*{.5in}
= \phi (p'_{\!\ssc (\!\varsigma\!)}- p_{\!\ssc (\!\varsigma\!)},
 x'_{\ssc (\!\varsigma\!)}-x_{\ssc (\!\varsigma\!)})
  e^{{i}(p_{\!\ssc (\!\varsigma\!)}x'_{\ssc (\!\varsigma\!)}- x_{\!\ssc (\!\varsigma\!)}p'_{\ssc (\!\varsigma\!)})}\;.
\eea
With the indices suppressed, the last expression has exactly 
the same form as in the $H\!(3)$ case, with the understanding 
that $p_{\!\ssc (\!\varsigma\!)}x'_{\ssc (\!\varsigma\!)}$, 
for example, stands for 
$p_{\!\ssc (\!\varsigma\!)}^a x'_{\ssc (\!\varsigma\!)a} 
 = \delta_{ab} p_{\!\ssc (\!\varsigma\!)}^a x'^b_{\ssc (\!\varsigma\!)}$.
The abstract formulation from the set of canonical 
coherent states based on the $H\!(4)$ manifold and the 
one from the irreducible component of the regular 
representation are hence really the same one. 

In correspondence with the case of $H\!(3)$,  we can take
$\left|0_{\ssc (\!\varsigma\!)}\rra$ as the zero eigenstate of 
$\hat{X}_{\ssc (\!\varsigma\!)a} + i \hat{P}_{\ssc (\!\varsigma\!)a}$ operator,
{\em i.e.} $\left( \hat{X}_{\ssc (\!\varsigma\!)a} + i \hat{P}_{\ssc (\!\varsigma\!)a} \right)
  \left|0_{\ssc (\!\varsigma\!)}\rra =0$, with its  wavefunction
\[
\phi_o(p_{\!\ssc (\!\varsigma\!)},x_{\ssc (\!\varsigma\!)}) 
  =\lla p_{\!\ssc (\!\varsigma\!)},x_{\ssc (\!\varsigma\!)}  |0,0\rra
=e^{-\frac{p_{\!\ssc (\!\varsigma\!)}^2+x_{\!\ssc (\!\varsigma\!)}^2}{2}}\;,
\]
the symmetric Gaussian. The function corresponds to 
the $n=0$ state wavefunction of the covariant harmonic 
oscillator and is actually invariant under the Lorentz 
transformations of $SO(1,3)$, which really has the 
boosts part nonunitarily represented, as to be explicitly 
shown below. The representation space spanned by the 
overcomplete set of coherent states is the same as that 
of the span of all the harmonic oscillator Fock states 
with the quite standard mathematical relationship
between the two sets of bases, which we present
in the appendix. The wavefunction for the
$\left| p_{\!\ssc (\!\varsigma\!)A},x_{\ssc (\!\varsigma\!)A} \rra$ 
state can be given by 
$\phi_{\!\ssc A}(p_{\!\ssc (\!\varsigma\!)},x_{\ssc (\!\varsigma\!)}) 
=  {U}^{\ssc\! L}_{\!\ssc \varsigma}\!( p_{\!\ssc (\!\varsigma\!)A},x_{\ssc (\!\varsigma\!)A}) 
  \phi_o (p_{\!\ssc (\!\varsigma\!)},x_{\ssc (\!\varsigma\!)})$.
Explicitly, we have 
\bea\label{aolap}
\phi_{\!\ssc A} \equiv
 \lla p^a_{\!\ssc (\!\varsigma\!)},x^a_{\ssc (\!\varsigma\!)}| p^a_{\!\ssc (\!\varsigma\!)A}, x^a_{\ssc (\!\varsigma\!)A} \rra 
=e^{i\left( x_{\!\ssc (\!\varsigma\!)a} p\vsia^a-p_{\!\ssc (\!\varsigma\!)a} x\vsia^a \right) }
e^{-\frac{1}{2}\left[\left(x\vsi-x\vsia \right)^2+\left(p\vsi-p\vsia  \right)^2 \right]} \;.
\eea 

\section{Explicit \boldmath $H_{\!\ssc R}(1,3)$ Picture \label{sec3}}
 Let us illustrate explicitly how the above $H_{\!\ssc R}(4)$
representation picture serves as a pseudo-unitary
irreducible representation of the $H_{\!\ssc R}(1,3)$
we are really after. The spinless pseudo-unitary
representation for the latter can be seen as built 
from the representations of $Y_\mu$ and $E_\mu$
with the $H_{\!\ssc R}(3)$ part, $Y_i$ and $E_i$, being
Hermitian. The matching to the unitary representation
of $H_{\!\ssc R}(4)$ is through
\[
Y_{\!\ssc 0} \leftrightarrow iY_{\!\ssc 4} \;, \qquad
E_{\!\ssc 0} \leftrightarrow iE_{\!\ssc 4} \;.
\]
That is to say,  $Y_{\!\ssc 0}$ and $E_{\!\ssc 0}$ are
represented explicitly by 
\begin{equation}\label{xp0}
\hat{X}^{\!\ssc L}_{\!\ssc (\!\varsigma\!)0}
\equiv i\hat{X}^{\!\ssc L}_{\!\ssc (\!\varsigma\!)4}\;,
\qquad 
\hat{P}^{\!\ssc L}_{\!\ssc (\!\varsigma\!)0}
\equiv i\hat{P}^{\!\ssc L}_{\!\ssc (\!\varsigma\!)4}\;.
\end{equation}
The relation $J_{\mu\nu}=Y_\mu  E_\nu - Y_\nu  E_\mu$ 
completes the representation. The representation space 
is essentially the same as the Hilbert space spanned 
by the Fock states of the corresponding covariant 
harmonic oscillator \cite{082}, or equivalently the space 
of the wavefunctions within the class of rapidly decreasing 
functions formulated as $\psi(x^a\vsi)$ in Ref.\cite{083}. 
Here we have a formulation of the wavefunctions based on 
the coherent state basis though, and a sketch of 
its relation to the Fock state basis  formulation is 
given in the appendix. 

The most important point to note is that the above
used inner product $\lla\cdot|\cdot\rra$, which for
the wavefunction representation corresponds to the 
usual squared-integral, though convenient to be used
in most of the analysis including the above basis
definition of the wavefunctions in relation to the 
abstract state vectors, is not the inner product of
physical interest. It is not preserved by the Lorentz
transformations. The physical, Lorentz invariant, 
inner product is given in terms of the parity operator 
$\mathcal{P}_{\!\ssc 4}$, which sends $x^{\ssc 4}\vsi$ 
to $-x^{\ssc 4}\vsi$ and $p^{\ssc 4}\vsi$ to $-p^{\ssc 4}\vsi$
 \cite{083}, as
\bea \label{P4}
\lla\!\lla \phi | \phi' \rra\!\rra
= \lla \phi | \mathcal{P}_{\!\ssc 4} | \phi' \rra 
=\frac{1}{\pi^4}  \int\!\! d^4p\vsi d^4x\vsi \; 
     \bar{\phi}(p_{\!\ssc (\!\varsigma\!)_i}, -p_{\!\ssc (\!\varsigma\!)_4},
     x_{\!\ssc (\!\varsigma\!)_i},-x_{\!\ssc (\!\varsigma\!)_4}) \, {\phi'}\!(p^a\vsi,x^a\vsi) \;.
\eea
In fact, all our $H_{\!\ssc R}(1,3)$ generators are 
pseudo-Hermitian with respect to 
$\mathcal{P}_{\!\ssc 4}$, \emph{i.e.}\footnote{
In particular, the operators corresponding to $Y_{\ssc \!0}$ 
and $E_{\ssc \!0}$, and therefore also to the Lorentz 
boost generators $J_{{\ssc 0}i}$, given by 
$i\hat{X}^{\!\ssc L}_{\!\ssc (\!\varsigma\!)4}$,
$i\hat{P}^{\!\ssc L}_{\!\ssc (\!\varsigma\!)4}$ and
$i\left(\hat{X}^{\!\ssc L}_{\!\ssc (\!\varsigma\!)4}\hat{P}^{\!\ssc L}_{{\!\ssc (\!\varsigma\!)}i}
  - \hat{X}^{\!\ssc L}_{{\!\ssc (\!\varsigma\!)}i}\hat{P}^{\!\ssc L}_{\!\ssc (\!\varsigma\!)4}\right)$, 
respectively, are anti-Hermitian satisfying 
$\hat{A}^\dag=-\hat{A}$, while the rest of the
generators are Hermitian and commute with 
$\mathcal{P}_{\!\ssc 4} =\mathcal{P}_{\!\ssc 4}^{-1}$.
}
\begin{equation} \label{pseud}
    \hat{A}^\dag=\mathcal{P}_{\!\ssc 4}  \hat{A} \mathcal{P}_{\!\ssc 4}^{-1} \;.
\end{equation}
For our Lorentz  invariant inner product, introduced 
first in Ref.\cite{083},  we take
$\left.\!\!\left| \phi \rra\!\rra \!\equiv \left| \phi \rra$,
and the new functional (bra) as
$\lla\!\lla \phi \right|\!\!\right. \equiv \!\lla \phi \right| \mathcal{P}_{\!\ssc 4}$.
The pseudo-Hermitian nature is exactly the self-adjointness 
with respect to the invariant inner product, {\em i.e.} 
$\langle\!\langle  \hat{A} \cdot | = \lla\!\lla  \cdot \right|\!\!\right. \hat{A}$~,
hence may be called $\mathcal{P}_{\!\ssc 4}$-Hermitian.
That is exactly in line with the more general studies 
of pseudo-Hermitian quantum mechanics  \cite{DM}\footnote{
Note that in the literature, the term $\mathcal{P}_{\!\ssc 4}$-pseudo-Hermitian
is commonly used for operators $\hat{A}$ satisfying (\ref{pseud}), 
while $\mathcal{P}_{\!\ssc 4}$-pseudo-unitary used for 
operators satisfying $V^{-1}=\mathcal{P}^{-1}_{\!\ssc 4} V^\dag\mathcal{P}_{\!\ssc 4}$, 
like those of the one-parameter group $V_{(s)}=e^{is\hat{A}}$.
}. 
Note that the usual studies of the latter focus on systems 
with pseudo-Hermitian physical Hamiltonians while we
are talking here about pseudo-Hermitian generators
of a pseudo-unitary representation of the background
(relativity) symmetry group without specifying a 
Hamiltonian. We want to emphasize that quantum
dynamics is symplectic dynamics and the  physical 
Hamiltonian is just a one among the many general
Hamiltonians with the generated Hamiltonian
flows as symmetries of the phase space. It is the 
symplectic structure of the latter as fixed by the 
invariant inner product that is really the key. 

Each $\mathcal{P}_{\!\ssc 4}$-Hermitian generator 
generates a  one-parameter group of transformations,
the $\mathcal{P}_{\!\ssc 4}$-unitary transformations 
which preserve the invariant inner product. In fact,
the latter was constructed from that requirement
 \cite{083}. It is important to note that the 
$\mathcal{P}_{\!\ssc 4}$-unitarity is certainly not
unitarity in any sense  as the inner product is not 
positive definite, which can be seen explicitly in the 
eigenstate basis of the covariant Harmonic oscillator
problem. In fact, exactly as in Minkowski spacetime, 
we have positive norm spacelike states, negative norm 
timelike states, as well as lightlike states with 
vanishing norm,  which is the basic feature
of the representation we want \cite{083,082}.

Normalization of states with respect to two
different inner products are of course different,
and our coherent states $\left|p^a\vsi,x^a\vsi \rra$
are only normalized with respect to the unphysical
inner product $\lla\cdot|\cdot\rra$. It is then easy
to see that
\bea
\lla\!\lla p^a\vsi,x\vsi^a|p^a\vsi,x^a\vsi \rra\!\rra
 = e^{-2\left(x\vsi^{\ssc 4}\!\right)^{\!2}-2\left(p\vsi^{\ssc 4}\!\right)^{\!2} } \;.
\eea

In fact, the true $H(1,3)$ coherent states can be introduced
within the same Hilbert space using the 
$\mathcal{P}_{\!\ssc 4}$-unitary operator representing
its element. They are given by 
\bea
\left.\left|p^\mu\vsi,x^\mu\vsi \rra\!\rra
\equiv V_{\ssc \!\varsigma}\!(p^\mu_{\!\ssc (\!\varsigma\!)},x^\mu_{\ssc (\!\varsigma\!)})
       \left.\left|0,0 \rra\!\rra
 \equiv e^{-i\theta_{\ssc (\!\varsigma\!)}} 
 {V}_{\ssc \!\varsigma}(p^\mu_{\!\ssc (\!\varsigma\!)},x^\mu_{\ssc (\!\varsigma\!)},
     \theta_{\!\ssc (\!\varsigma\!)})\left.\left|0,0 \rra\!\rra  \;, 
\eea
with $\left.\left|0,0 \rra\!\rra = \left|0_{\ssc (\!\varsigma\!)}\rra$, 
where
\bea
V_{\ssc \!\varsigma}(p^\mu_{\!\ssc (\!\varsigma\!)},x^\mu_{\ssc (\!\varsigma\!)},\theta_{\ssc (\!\varsigma\!)})
 \equiv e^{i (p^\mu_{\!\ssc (\!\varsigma\!)}\hat{X}_{\ssc (\!\varsigma\!)\mu}
  - x^\mu_{\ssc (\!\varsigma\!)}\hat{P}_{\ssc (\!\varsigma\!)\mu} 
     +\theta_{\!\ssc (\!\varsigma\!)}\hat{I})} \;,
\eea
represents the group element
$\widetilde{W}\!(p^\mu\vsi, x^\mu\vsi, \theta\vsi)$ satisfying
\bea\label{wwt}
\widetilde{W}\!(p'^\mu\vsi, x'^\mu\vsi, \theta'\vsi)  \widetilde{W}\!(p^\mu\vsi, x^\mu\vsi, \theta\vsi)
= \widetilde{W}\!\!\left(p'^\mu\vsi + p^\mu\vsi, x'^\mu\vsi + x^\mu\vsi, 
     \theta'\vsi+\theta\vsi- \!(x'_{{\ssc (\!\varsigma\!)}_\mu}  p^\mu\vsi \!- p'_{{\ssc (\!\varsigma\!)}_\mu} x^\mu\vsi) \right) .
\eea 
As $V_{\ssc \!\varsigma}\!(p^\mu\vsi,x^\mu\vsi)$
preserves the $\lla\!\lla \cdot| \cdot \rra\!\rra$
inner product, we have 
${\lla\!\!\lla p^\mu\vsi,x\vsi^\mu| p^\mu\vsi,x^\mu\vsi \rra\!\!\rra } =1$,
hiding the nature of the norm as non-positive definite, 
in the same way as all the 
$\lla\!\!\lla p^a\vsi,x\vsi^a|p^a\vsi,x^a\vsi \rra\!\!\rra$ 
norms are positive. As given in Ref.\cite{083} and summarized 
in the appendix, the orthonormal basis states of the Fock space,
based on which the invariant inner product and the 
$\mathcal{P}_{\!\ssc 4}$ operator were first defined,
satisfy $\lla\!\lla m| n\rra\!\rra = (-1)^{n_{ \!4}} \delta_{mn}$.
Actually, we have just the usual coherent state representation, with
\bea &&
{\lla\!\!\lla p^\mu\vsi,x\vsi^\mu| \hat{X}_\nu |p^\mu\vsi,x^\mu\vsi \rra\!\!\rra }
=2x_\nu \;,
\sea
{\lla\!\!\lla p^\mu\vsi,x\vsi^\mu| \hat{P}_\nu |p^\mu\vsi,x^\mu\vsi \rra\!\!\rra }
=2p_\nu \;,
\eea
which would naively be thought of as being unitary.
The $\mathcal{P}_{\!\ssc 4}$-Hermitian nature of
$\hat{X}_\mu$ and $\hat{P}_\mu$ and
$\mathcal{P}_{\!\ssc 4}$-unitary nature of
$V_{\ssc \!\varsigma}\!(p^\mu\vsi,x^\mu\vsi)$
are completely hidden. They look as good as
Hermitian and unitary in the naive sense.  

The wavefunctions in this basis can  be introduced as  
$\tilde\phi(p^\mu\vsi,x\vsi^\mu) \equiv \lla\!\!\lla p^\mu\vsi,x\vsi^\mu| \phi\rra\!\!\rra$ 
on which we have again 
\bea 
 \lla\!\!\lla p^\mu_{\!\ssc (\!\varsigma\!)},x^\mu_{\ssc (\!\varsigma\!)}  
  \left| \hat{X}_{\!\ssc (\!\varsigma\!)\nu} \right|\phi \rra\!\!\rra 
  &=& \hat{X}^{\!\ssc \tilde{L}}_{\!\ssc (\!\varsigma\!)\nu} 
     \tilde\phi (p^\mu_{\!\ssc (\!\varsigma\!)},x^\mu_{\ssc (\!\varsigma\!)})  \;,
\nonumber \\
\lla\!\!\lla p^\mu_{\!\ssc (\!\varsigma\!)},x^\mu_{\ssc (\!\varsigma\!)}  
  \left| \hat{P}_{\!\ssc (\!\varsigma\!)\nu} \right|\phi \rra\!\!\rra 
   &=& \hat{P}^{\!\ssc \tilde{L}}_{\!\ssc (\!\varsigma\!)\nu} 
   \tilde\phi (p^\mu_{\!\ssc (\!\varsigma\!)},x^\mu_{\ssc (\!\varsigma\!)}) \;,
\eea
with
\bea
\hat{X}^{\!\ssc \tilde{L}}_{\!\ssc (\!\varsigma\!)\mu} &=& 
   x_{\ssc (\!\varsigma\!)\mu}  +  i \partial_{\!p_{\!\ssc (\!\varsigma\!)}^\mu }\;,
\nonumber \\
\hat{P}^{\!\ssc \tilde{L}}_{\!\ssc (\!\varsigma\!)\mu} &=& 
   p_{\ssc (\!\varsigma\!)\mu}  - i \partial_{\!x_{\!\ssc (\!\varsigma\!)}^\mu }\;,
\label{L}
\eea
and
\bea \label{v-shift} &&
{V}^{\ssc\! \tilde{L}}_{\ssc \!\varsigma}\!(p^\mu_{\!\ssc (\!\varsigma\!)},x^\mu_{\ssc (\!\varsigma\!)}) 
  \tilde\phi(p'^\mu_{\!\ssc (\!\varsigma\!)},x'^\mu_{\ssc (\!\varsigma\!)})   \equiv
 \lla\!\!\lla p'^\mu_{\!\ssc (\!\varsigma\!)},x'^\mu_{\ssc (\!\varsigma\!)} 
 \left|{V}_{\ssc \!\varsigma}(p^\mu_{\!\ssc (\!\varsigma\!)},x^\mu_{\ssc (\!\varsigma\!)})  \right|\phi \rra\!\!\rra 
\sea\hspace*{.5in}
= \tilde\phi (p'^\mu_{\!\ssc (\!\varsigma\!)}- p^\mu_{\!\ssc (\!\varsigma\!)},
 x'^\mu_{\ssc (\!\varsigma\!)}-x^\mu_{\ssc (\!\varsigma\!)})
  e^{{i}(p^\mu_{\!\ssc (\!\varsigma\!)}x'_{\ssc (\!\varsigma\!)\mu}- x^\mu_{\!\ssc (\!\varsigma\!)}p'_{\ssc (\!\varsigma\!)\mu})}\;.
\eea
One can also write 
$\hat{X}^{\!\ssc \tilde{L}}_{\!\ssc (\!\varsigma\!)\mu}=x_{\!\ssc (\!\varsigma\!)\mu}\star$
and $\hat{P}^{\!\ssc \tilde{L}}_{\!\ssc (\!\varsigma\!)\mu}=p_{\!\ssc (\!\varsigma\!)\mu}\star$
with the Moyal star product for the Minkowski four vectors. 
We have also
\bea\label{molap}
\tilde\phi_{\!\ssc A} (p^\mu\vsi,x^\mu\vsi)\equiv
 \lla\!\! \lla p^\mu\vsi,x^\mu\vsi| p^\mu_{\!\ssc (\!\varsigma\!)A}, x^\mu_{\ssc (\!\varsigma\!)A} \rra\!\!\rra
=e^{i\left( x_{\!\ssc (\!\varsigma\!)\mu} p\vsia^\mu-p_{\!\ssc (\!\varsigma\!)\mu} x\vsia^\mu \right) }
e^{-\frac{1}{2}\left[\left(x\vsi-x\vsia \right)^2+\left(p\vsi-p\vsia  \right)^2 \right]} \;,
\eea
where $\left(x\vsi-x\vsia \right)^2$ and
$\left(p\vsi-p\vsia  \right)^2$ here are the Minkowski 
vector magnitude squares. It is important to distinguish 
$\left.\left|p^\mu\vsi,x^\mu\vsi \rra\!\!\rra$ 
from  $\left.\left|p^a\vsi,x^a\vsi \rra\!\!\rra$ and 
$\tilde\phi (p^\mu_{\!\ssc (\!\varsigma\!)},x^\mu_{\ssc (\!\varsigma\!)})$ 
from $\phi (p^a_{\!\ssc (\!\varsigma\!)},x^a_{\ssc (\!\varsigma\!)})$,
the relations between which are not easy to see from the
results here.  However, based on the  analysis in the Fock 
state basis, given in the appendix, they can easily be 
understood. Note that ${V}_{\ssc \!\varsigma}\!(p^\mu\vsi,x^\mu\vsi)$
and ${U}_{\ssc \!\varsigma}\!(p^a\vsi,x^a\vsi)$
cannot be identified for all real $p^\mu\vsi$, $x^\mu\vsi$,
$p^a\vsi$,  and $x^a\vsi$, just for the trivial
case with nonzero only  $p^i\vsi$ and $x^i\vsi$.
Otherwise, ${V}_{\ssc \!\varsigma}\!(p^\mu\vsi,x^\mu\vsi)$
is not unitary and ${U}_{\ssc \!\varsigma}\!(p^a\vsi,x^a\vsi)$
not $\mathcal{P}_{\!\ssc 4}$-unitary. We are
not interested in imaginary values of the parameters.

The representation based on 
$\hat{X}^{\!\ssc \tilde{L}}_{\!\ssc (\!\varsigma\!)\mu}$
and $\hat{P}^{\!\ssc \tilde{L}}_{\!\ssc (\!\varsigma\!)\mu}$
can obviously be obtained as an irreducible component 
of the regular representation of $H(1,3)$, seen as 
a subgroup of $H_{\!\ssc R}(1,3)$, along the same 
line as described for the  $H_{\!\ssc R}(4)$ case
in the previous section. However, a naive analysis 
of the formulation along that line would completely 
hide the pseudo-unitary nature of the representation 
and give the $\tilde\phi_n (p^\mu\vsi,x^\mu\vsi)$
wavefunctions of the Fock states as having various 
divergence issues  \cite{083,Z}, as well as suggest 
restriction to the spacelike or timelike domains of 
the variables. The latter is not compatible with the 
coherent state picture itself. Here, the problems are 
resolved, and all $\tilde\phi_n (p^\mu\vsi,x^\mu\vsi)$
without restricting the domain have the proper norm 
$\pm 1$. To illustrate the feature
explicitly, we first note that from the analysis
in the appendix, we have the identification of
$\left.\left|p^\mu\vsi,x^\mu\vsi \rra\!\!\rra$ 
with $e^{\left(x\vsi^{\ssc 4}\!\right)^{\!2}+\left(p\vsi^{\ssc 4}\!\right)^{\!2} }  
         \left.\left|p^a\vsi,x^a\vsi \rra\!\!\rra$,
under the parameter relations 
$x^{\ssc 0}=p^{\ssc 4}$ and $p^{\ssc 0}=-x^{\ssc 4}$.
That gives the resolution of the identity
\bea
\hat{I} &=&\int\! \frac{d^3\!p\vsi d^3\!x\vsi dp^{\ssc 4}\vsi  dx^{\ssc 4}\vsi}{\pi^4}
        \left|p^a\vsi,x^a\vsi \rra \! \lla p^a\vsi,x^a\vsi \right|
\sea
   = \int \!\! {d^3\!p\vsi d^3\!x\vsi dp^{\ssc 0}\vsi  dx^{\ssc 0}\vsi}
    \frac{e^{-2\left(x\vsi^{\ssc 0}\!\right)^{\!2}-2\left(p\vsi^{\ssc 0}\!\right)^{\!2} }  }{\pi^4}
   \left.\left|p^\mu\vsi,x^\mu\vsi \rra\!\!\rra \! \lla\!\!\lla p^\mu\vsi,x\vsi^\mu \right|\right. 
    \mathcal{P}_{\!\ssc 4} \;.
\eea
We have hence the functional  $\lla\!\lla \psi \right|\right.$
represented on the space of $\tilde\phi (p^\mu\vsi,x^\mu\vsi)$ as
\[
\int \!\! {d^3\!p\vsi d^3\!x\vsi dp^{\ssc 0}\vsi  dx^{\ssc 0}\vsi}
    \frac{e^{-2\left(x\vsi^{\ssc 0}\!\right)^{\!2}-2\left(p\vsi^{\ssc 0}\!\right)^{\!2} }  }{\pi^4}
   \tilde\psi^*\! (p^i\vsi,x^i\vsi, -p\vsi^{\ssc 0},-x\vsi^{\ssc 0}) \,\bigg(\cdot\bigg) \;,
\]
with the very nontrivial integration measure.
The inner product $\lla\!\lla \psi | \phi \rra\!\rra$ 
is then given by
\bea \label{ptip}
  \frac{1}{\pi^4}\int \!\! {d^3\!p\vsi d^3\!x\vsi dp^{\ssc 0}\vsi  dx^{\ssc 0}\vsi}
   \frac{\tilde\psi^*\! (p^i\vsi,x^i\vsi, -p\vsi^{\ssc 0},-x\vsi^{\ssc 0}) }{e^{\left(x\vsi^{\ssc 0}\!\right)^{\!2}+\left(p\vsi^{\ssc 0}\!\right)^{\!2} }  } 
  \frac{\tilde\phi (p^\mu\vsi,x^\mu\vsi)}{e^{\left(x\vsi^{\ssc 0}\!\right)^{\!2}+\left(p\vsi^{\ssc 0}\!\right)^{\!2} } }\;.
\eea
Each of the basis functions ${\tilde\phi_n (p^\mu\vsi,x^\mu\vsi)}$,
and hence any general ${\tilde\phi (p^\mu\vsi,x^\mu\vsi)}$ 
in the spanned space, is formally divergent at timelike 
infinity of the four-vector variables. On the other hand, all
$\frac{\tilde\phi_n (p^\mu\vsi,x^\mu\vsi)}{e^{\left(x\vsi^{\ssc 0}\!\right)^{\!2}+\left(p\vsi^{\ssc 0}\!\right)^{\!2} } }$,
and hence all 
$\frac{\tilde\phi (p^\mu\vsi,x^\mu\vsi)}{e^{\left(x\vsi^{\ssc 0}\!\right)^{\!2}+\left(p\vsi^{\ssc 0}\!\right)^{\!2} } }$,
are rapidly decreasing functions like the corresponding
 ${\phi_n (p^a\vsi,x^a\vsi)}$ and  ${\phi (p^a\vsi,x^a\vsi)}$.
The factor ${e^{-\left(x\vsi^{\ssc 0}\!\right)^{\!2}-\left(p\vsi^{\ssc 0}\!\right)^{\!2} } }$
 takes the ${e^\frac{{\left(x\vsi^{\ssc 0}\!\right)^{\!2}+\left(p\vsi^{\ssc 0}\!\right)^{\!2} } }{2} }$
factor in all ${\tilde\phi_n (p^\mu\vsi,x^\mu\vsi)}$ back to
${e^-\frac{{\left(x\vsi^{\ssc 0}\!\right)^{\!2}+\left(p\vsi^{\ssc 0}\!\right)^{\!2} } }{2} }$, which
characterizes the class of functions. The integral is
finite for all wavefunctions as finite linear combinations
of the Fock state basis $\tilde{\phi}_n$.  Using
$\frac{\tilde\phi (p^\mu\vsi,x^\mu\vsi)}{e^{\left(x\vsi^{\ssc 0}\!\right)^{\!2}+\left(p\vsi^{\ssc 0}\!\right)^{\!2} } }$
as the wavefunctions cannot be correct, though.
That would, for example, make the wavefunction for 
$\left.\left|0 \rra\!\rra$ not Lorentz invariant and 
mess up the right transformation properties of all those
for the Fock states, described in Ref.\cite{083}.
Thinking further about 
${\tilde\psi^*\! (p^i\vsi,x^i\vsi, -p\vsi^{\ssc 0},-x\vsi^{\ssc 0}) }$
as ${\tilde\psi^*\! (p_{\!\ssc (\!\varsigma\!)\mu},x_{\!\ssc (\!\varsigma\!)\mu}) }$,
one can see in hindsight that the inner 
product expression is indeed exactly what it should be.
Of course we have that here rigorously established.

\section{Lorentz to Galilean Contraction} \label{sec4}
A contraction of the Lorentz symmetry $SO(1,3)$, sitting 
inside the $H_{\!\ssc R}(1,3)$, to the Galilean $ISO(3)$ 
has been discussed in Ref.\cite{071}, together with the 
corresponding coset spaces of interest. The full (quantum)
relativity symmetry group obtained by contraction is named 
$H_{\!\ssc GH}(3)$, with commutators among generators 
essentially given by
\bea &&
[J_{ij}, J_{hk}] =2i  (\delta_{jk}J_{ih}+  \delta_{ih}J_{jk}  -\delta_{ik}J_{jh} - \delta_{jh}J_{ik}) \;,
\sea
[J_{ij}, X_k] = -2i(\delta_{jk} X_i - \delta_{ik} X_j) \;,
\qquad
[J_{ij}, P_k] = -2i(\delta_{jk} P_i - \delta_{ik} P_j) \;,
\sea
[J_{ij}, K_k] = -2i(\delta_{jk} K_i - \delta_{ik} K_j) \;,
\qquad
[K_i, K_j] = 
 0 \;,
\nonumber \\ &&
[K_i, H] = 2iP_i \;,
\qquad
[K_i, P_j] = 
 0 \;,
\quad\quad
[X_i,P_j]= 2i\delta_{ij} I' \;,
\nonumber \\ &&
[T , H]= -2i I' \;,
\quad\quad
[K_i, T ] =
 0 \;,
\quad\quad
[K_i, X_j]=  2i\delta_{ij}  T \;.
\eea
Note that the full result for the other commutators beyond the $J_{ij}$
and $K_i$ set, originated from $SO(1,3)$, is essentially fixed by the
requirement of having the Galilean $K_i$-$H$ and the Heisenberg $X$-$P$
commutators. However, for the purpose here, the explicit contraction is 
to be implemented a bit differently. It is taken as the $c \to \infty$ 
limit of $K_i= \frac{1}{c} J_{{\ssc 0}i}$,  $P_i = \frac{1}{c} E_i$,  
$X_i=\frac{1}{c} Y_i$,$T =\frac{-1}{c^2} Y_{\ssc 0}$, $I'=\frac{1}{c^2} I$, 
with the renaming $H\equiv -E_{\ssc 0}$.  In the contraction, 
$K_i$ as generators for the Galilean boosts are the basic starting 
point and we would like to be able to trace physics, including the 
relative physical dimensions of quantities, by considering the 
speed of light $c$ as having a physical dimension. Introducing 
$X_i=\frac{1}{c} Y_i$ is to keep the same physical dimensions 
for $X_i$ and $P_i$. However, the essence of the contraction 
scheme as a formulation to retrieve an approximate physical 
theory from a more exact one is really to implement the 
contraction at a representation level.

To implement the contraction on a ${U}^{\ssc\! L}_{\!\ssc \varsigma}$, 
or the matching ${U}_{\!\ssc \varsigma}$ as a representation 
of the original $H\!(1,3)$, it is important to note that the original 
central charge generator $I$ represented by $\varsigma \hat{I}$
in ${U}_{\!\ssc \varsigma}$ would give the representation of the 
contracted $I'$, which remains central, as $\frac{\varsigma}{c^2} \hat{I}$. 
For a sensible result, one needs to consider $\varsigma=c^2  \chi$ 
with $\chi$ staying finite at the contraction limit, hence $I'$ 
represented by  $\chi \hat{I}$ (recall: $\hat{I}$ is the identity 
operator). Hence,  ${U}_{\!\ssc \varsigma}$ contracts into  
${U}_{\!\ssc \chi}$. In another words, the ${U}_{\!\ssc \varsigma}$ 
representation of the original $H\!(1,3)$, and the full 
$H_{\!\ssc  R}(1,3)$, survives as the ${U}_{\!\ssc \chi}$ 
($\chi=\frac{\varsigma}{c^2}> 0$) representation of the 
$H\!(3)$ in the contracted $H_{\!\ssc G\!H}(3)$, as well 
as of the full group.

With the results from the last section, however, 
we can and prefer to work on the equivalent
${V}^{\ssc\! \tilde{L}}_{\!\ssc \varsigma}$ representation, 
as well as using ${V}_{\!\ssc \varsigma}$, instead of  
${U}_{\!\ssc \varsigma}$, and the 
$\left.\!\left|p\vsi^\mu,x\vsi^\mu \rra\!\!\rra$ basis. 
In any case, all representations should be taken with 
the  physical invariant inner product. We have 
the exact parallel of ${V}_{\!\ssc \varsigma}$ contracting to
${V}_{\!\ssc \chi}$. For the $c \to \infty$ limit of
${V}^{\ssc\! \tilde{L}}_{\!\ssc \chi}\!(p\vsi^\mu,x\vsi^\mu)$,  
we have to consider first
\[  
\hat{P}^{\ssc\! \tilde{L}}_{\!\ssc \chi i} = \frac{1}{c} \hat{E}^{\ssc\! \tilde{L}}_{\varsigma i} \;, 
\quad 
\hat{X}^{\ssc\! \tilde{L}}_{\!\ssc \chi i} = \frac{1}{c} \hat{Y}^{\ssc\! \tilde{L}}_{\varsigma i}\;, 
 \quad
\hat{H}^{\ssc\! \tilde{L}}_{\!\ssc \chi} =  -\hat{E}^{\ssc\! \tilde{L}}_{\varsigma\ssc 0}\;,
\quad
\hat{T}^{\ssc\! \tilde{L}}_{\!\ssc \chi} = -\frac{1}{c^2} \hat{Y}^{\ssc\! \tilde{L}}_{\varsigma \ssc 0} \;,
\] 
and take that to obtain
\bea && \label{resc}
\hat{X}^{\ssc\! \tilde{L}}_{\!\ssc (\!\chi\!) i} = \frac{1}{\sqrt{\chi}} \hat{X}^{\ssc\! \tilde{L}}_{\!\ssc \chi i}
= \hat{X}^{\ssc\! \tilde{L}}_{(\!\varsigma\!)i} \;,
\qquad
\hat{P}^{\ssc\! \tilde{L}}_{\!\ssc (\!\chi\!) i} = \frac{1}{\sqrt{\chi}} \hat{P}^{\ssc\! \tilde{L}}_{\!\ssc \chi i}
= \hat{P}^{\ssc\! \tilde{L}}_{(\!\varsigma\!)i} \;,
\sea
\hat{T}^{\ssc\! \tilde{L}}_{\!\ssc (\!\chi\!)} = \frac{1}{\sqrt{\chi}} \hat{T}^{\ssc\! \tilde{L}}_{\!\ssc \chi}
= -\frac{1}{c} \hat{X}^{\ssc\! \tilde{L}}_{(\!\varsigma\!)\ssc 0} \;,
\qquad
\hat{H}^{\ssc\! \tilde{L}}_{\!\ssc (\!\chi\!)} = \frac{1}{\sqrt{\chi}} \hat{H}^{\ssc\! \tilde{L}}_{\!\ssc \chi}
=- c \hat{P}^{\ssc\! \tilde{L}}_{(\!\varsigma\!)\ssc 0} \;,
\eea
(with $\varsigma = c^2 \chi$). The above are 
the basic set of operators acting on the functional space of 
 $\phi (p_{\!\ssc (\!\varsigma\!)},x_{\ssc (\!\varsigma\!)})$, 
with the variables properly rescaled to a new set of
variables to match with the operators. There is also the exactly 
corresponding set of operators, $\hat{X}_{{\!\ssc (\!\chi\!)}i}$,
$\hat{P}_{{\!\ssc (\!\chi\!)}i}$, $\hat{T}_{{\!\ssc (\!\chi\!)}}$,
and $\hat{H}_{{\!\ssc (\!\chi\!)}}$, and ${V}_{\!\ssc \chi}$
on the abstract Hilbert space which are helpful for tracing
 the proper description. The proper labels for the states  
$\left.\!\left|p\vsi^\mu,x\vsi^\mu \rra\!\!\rra$
at the contraction limit should be 
$\left.\!\left| p^i_{\!\ssc (\!\chi\!)}, {e}_{\!\ssc (\!\chi\!)}, 
      x^i_{\!\ssc (\!\chi\!)}, {t}_{\!\ssc (\!\chi\!)}  \rra\!\!\rra$,
satisfying
\bea &&
2 x _{{\!\ssc (\!\chi\!)}i} = 
  \lla\!\!\lla p^i_{\!\ssc (\!\chi\!)}, {e}_{\!\ssc (\!\chi\!)},   x^i_{\!\ssc (\!\chi\!)}, {t}_{\!\ssc (\!\chi\!)}    
      \left| \hat{X}_{{\!\ssc (\!\chi\!)}i} \right| 
    p^i_{\!\ssc (\!\chi\!)}, {e}_{\!\ssc (\!\chi\!)},   x^i_{\!\ssc (\!\chi\!)}, {t}_{\!\ssc (\!\chi\!)} \rra\!\!\rra \;,
    \sea
2 p _{{\!\ssc (\!\chi\!)}i} = 
  \lla\!\!\lla  p^i_{\!\ssc (\!\chi\!)}, {e}_{\!\ssc (\!\chi\!)},   x^i_{\!\ssc (\!\chi\!)}, {t}_{\!\ssc (\!\chi\!)} 
         \left| \hat{P}_{{\!\ssc (\!\chi\!)}i} \right| 
   p^i_{\!\ssc (\!\chi\!)}, {e}_{\!\ssc (\!\chi\!)},   x^i_{\!\ssc (\!\chi\!)}, {t}_{\!\ssc (\!\chi\!)}  \rra\!\!\rra \;,
\sea
2{t} _{\!\ssc (\!\chi\!)} = 
  \lla\!\!\lla p^i_{\!\ssc (\!\chi\!)}, {e}_{\!\ssc (\!\chi\!)},   x^i_{\!\ssc (\!\chi\!)}, {t}_{\!\ssc (\!\chi\!)}    
      \left| \hat{T}_{\!\ssc (\!\chi\!)} \right| 
   p^i_{\!\ssc (\!\chi\!)}, {e}_{\!\ssc (\!\chi\!)},   x^i_{\!\ssc (\!\chi\!)}, {t}_{\!\ssc (\!\chi\!)} \rra\!\!\rra \;,
\sea
2 {e} _{\!\ssc (\!\chi\!)} = 
  \lla\!\!\lla p^i_{\!\ssc (\!\chi\!)}, {e}_{\!\ssc (\!\chi\!)},   x^i_{\!\ssc (\!\chi\!)}, {t}_{\!\ssc (\!\chi\!)}       
   \left| \hat{H}_{\!\ssc (\!\chi\!)} \right| 
   p^i_{\!\ssc (\!\chi\!)}, {e}_{\!\ssc (\!\chi\!)},  x^i_{\!\ssc (\!\chi\!)}, {t}_{\!\ssc (\!\chi\!)} \rra\!\!\rra \;,
\eea
and hence giving
\[
\tilde\phi(p\vsi^\mu,x\vsi^\mu)  
\quad \longrightarrow  \quad
\tilde\phi( p_{\!\ssc (\!\chi\!)}^i,{e}_{\!\ssc (\!\chi\!)},
   x_{\ssc (\!\chi\!)}^i,{t}_{\ssc (\!\chi\!)}\!)
\]
with 
\begin{align}\label{newpar}
    x _{\!\ssc (\!\chi\!)i} &= x _{\!\ssc (\!\chi\!)}^i = x_{\ssc (\!\varsigma\!)}^i \;,
&p _{\!\ssc (\!\chi\!)i} = p _{\!\ssc (\!\chi\!)}^i = p_{\ssc (\!\varsigma\!)}^i \;, &
\nonumber\\
{t} _{\!\ssc (\!\chi\!)} &=  \frac{1}{c}\, x_{\ssc (\!\varsigma\!)}^{\ssc 0} \;,
&{e}_{\!\ssc (\!\chi\!)} = c\, p_{\ssc (\!\varsigma\!)}^{\ssc 0} \;. &
\end{align} 
 We have then, at least formally,
\bea &&
\hat{X}^{\ssc\! \tilde{L}}_{\!\ssc (\!\chi\!)} = x_{\!\ssc (\!\chi\!)} + i \partial_{p_{\!\ssc (\!\chi\!)}} \;,
\qquad
\hat{P}^{\ssc\! \tilde{L}}_{\!\ssc (\!\chi\!)} = p_{\!\ssc (\!\chi\!)} - i \partial_{x_{\!\ssc (\!\chi\!)}} \;,
\sea
\hat{T}^{\ssc\! \tilde{L}}_{\!\ssc (\!\chi\!)} =  {t}_{\!\ssc (\!\chi\!)} - i \partial_{{e}_{\!\ssc (\!\chi\!)}} \;,
\qquad \;\,
\hat{H}^{\ssc\! \tilde{L}}_{\!\ssc (\!\chi\!)}  = {e}_{\!\ssc (\!\chi\!)} + i  \partial_{{t}_{\!\ssc (\!\chi\!)}} \;.
\label{XPTH}
\eea 
The crucial quantity controlling the nature of the representation is
\[  
\lla\!\!\lla p^i_{\!\ssc (\!\chi\!)B}, {e}_{\!\ssc (\!\chi\!)B},   x^i_{\!\ssc (\!\chi\!)B}, {t}_{\!\ssc (\!\chi\!)B}    |
   p^i_{\!\ssc (\!\chi\!)A}, {e}_{\!\ssc (\!\chi\!)A},   x^i_{\!\ssc (\!\chi\!)A}, {t}_{\!\ssc (\!\chi\!)A} \rra\!\!\rra \;.
\]
From the original
$\lla\!\!\lla p^\mu_{\!\ssc (\!\chi\!)B},x^\mu_{\ssc (\!\chi\!)B}|    
    p^\mu_{\!\ssc (\!\chi\!)A}, x^\mu_{\ssc (\!\chi\!)A} \rra\!\!\rra$, given in Eq.(\ref{molap}), we have it as
\[
e^{i \left(  {e}\shib {t}\shia -t\shib {e}\shia
    +\delta_{ij}  x\shib^{i }p^j\shia-\delta_{ij} p^{i}\shib x\shia^j \right) }
   e^{-\frac{1}{2}\left[\left(x^{i}\shib-x^{i}\shia \right)^2 
 - c^2\left( {t}\shib-{t}\shia \right)^2 +\left(p^{i}\shib-p^{i}\shia \right)^2 
    -\frac{1}{c^2}\left({e}\shib-{e}\shia  \right)^2\right]}
\]
to be taken at the  $c \to \infty$ limit. It holds
$e^{\frac{1}{2c^2}\left({e}\shib-{e}\shia  \right)^2} \rightarrow 1$,
but the $e^{\frac{c^2}{2}\left({t}\shib-\tilde{t}\shia  \right)^2}$ 
factor diverges in the limit, except for ${t}\shib={t}\shia$, which 
indicates that we should consider only the latter case. The magnitude 
of the overlap being independent of ${e}\shib$ and ${e}\shia$ is still 
puzzling. The answer to that comes from a more careful thinking about 
the nature of the variables ${e}\shi$. Unlike ${t}\shi = \frac{x^0\vsi}{c}$, 
which is to be taken to be finite as in the general spirit of 
symmetry contraction, ${e}\shi = c p^{\ssc 0}$ is of quite 
different nature. The Lie algebra contraction to begin with 
only has a relabeling $H= - E_{\ssc 0}$  involving no $c$.
One may wonder if the $c$ in  
$\hat{H}^{\ssc\! \tilde{L}}_{\!\ssc (\!\chi\!)}
      =- c \hat{P}^{\ssc\! \tilde{L}}_{(\!\varsigma\!)\ssc 0}$
should be taken as giving a diverging energy observable 
$\hat{H}^{\ssc\! \tilde{L}}_{\!\ssc (\!\chi\!)}$ for any
finite $\hat{P}^{\ssc\! \tilde{L}}_{(\!\varsigma\!)\ssc 0}$.
Furthermore, for an Einstein particle of the rest mass $m$, 
\[
e = m c^2 + \frac{p^ip_i}{2m} + \cdots 
\]
where the neglected terms involve negative powers of $c^2$.
At the $c \to \infty$ limit, it is indeed diverging. Even
$p^{\ssc 0}$ is diverging. That is the result of the rest
mass as an energy. Hence, it sure suggests that we should
take our variable ${e}\shi$ as infinite, and the 
`non-relativistic' energy we are interested in is the 
kinetic energy $\frac{p^ip_i}{2m}$ given by the limit of
$e - m c^2$. Taking that feature into our consideration, 
the Hilbert space of interest under the contraction is 
really only the space spanned by the $H(3)$ coherent states 
$\left.\!\left| p^i_{\!\ssc (\!\chi\!)},  x^i_{\!\ssc (\!\chi\!)} \rra\!\rra$
for a fixed time $t\shi$ and a formally infinite $e\shi$.
To be exact, we should be implementing that logic from
an Einstein particle to our quantum observables 
$\hat{H}_{\ssc (\!\chi\!)}$,  $\hat{P}_{\ssc (\!\chi\!)}^{\ssc 0}$,
and $\hat{P}_{\ssc (\!\chi\!)i}$ or their expectation
values, but the conclusion is the same. Readers will see 
below in our analysis of the dynamics that we naturally
have an extended situation with an admissible interaction
potential or potential energy, and the Einstein particle
corresponds to the case where the latter vanishes. Any finite 
potential energy obviously does not change the story here.
The coherent state wavefunction
  $\tilde\phi_{\!\ssc A} (p^\mu\vsi,x^\mu\vsi)$ is equal to
$\lla\!\lla p^\mu_{\!\ssc (\!\varsigma\!)},x^\mu_{\ssc (\!\varsigma\!)}| p^\mu_{\!\ssc (\!\varsigma\!)A}, x^\mu_{\ssc (\!\varsigma\!)A} \rra\!\rra$,
hence at the contraction limit there is no more dependence
on $t\shi$ and  $e\shi$ reducing it  essentially to just
$\tilde\phi_{\!\ssc A} (p^i\shi,x^i\shi)$ . The operator
$\hat{T}^{\ssc\! \tilde{L}}_{\!\ssc (\!\chi\!)}$ acts on the 
Hilbert space of wavefunctions  only as a multiplication by 
$t\shi$ and is just like classical, while 
$\hat{H}^{\ssc\! \tilde{L}}_{\!\ssc (\!\chi\!)}$ is not
physically relevant. Note that the full contracted 
representation is then simply unitary. The part of the
inner product $\lla\!\lla \cdots | \cdots \rra\!\rra$ 
independent of $p^{\ssc 0}\vsi$ and $x^{\ssc 0}\vsi$, 
hence $t\shi$ and $e\shi$, is exactly the usual one.

\section{Group Theoretically Based WWGM Framework with
Wavefunctions in Coherent State Basis} \label{sec5}
The above analysis gives a successful picture of
the phase space of the  $H_{\!\ssc R}(1,3)$ theory, 
giving in the Galilean limit the phase space of the
$H_{\!\ssc R}(3)$ theory at each fixed `time' value.
The phase spaces, or more exactly the corresponding
projective Hilbert spaces, are to be seen as the 
quantum models of the spacetime \cite{066,070,082}. 
The infinite dimensional manifolds give, at the proper
relativity symmetry contraction limit, the familiar 
finite dimensional classical models as approximation. 
The explicit results of the classical limit for the 
present case is presented in the section below. The 
merit of our group theoretical approach is that it 
gives a full dynamical theory associated with the 
corresponding spacetime model for each relativity 
symmetry, mutually connected through the
contraction/deformation pattern. The dynamical
theory is naturally a Hamiltonian theory from the 
symmetry of the phase space as symplectic geometry, 
while the phase space is the space(time) at the 
quantum level, splitting into the (configuration)
space(time) and momentum space only at the classical
limit, with the Heisenberg commutator trivialized.
The dynamics is better described on the algebra
of observables as the matching representation 
of the group C$^*$-algebra \cite{070}, which for 
a quantum theory can be seen as a noncommutative 
geometric picture of the phase space with the 
position and momentum operators as coordinates, 
otherwise identified as the infinite dimensional 
(projective) Hilbert space \cite{078,081}. 

\subsection{The Algebra of Observables, Symmetries, and Dynamics}
The algebra of observables is depicted essentially as 
the one from a WWGM formalism, as functions and 
distributions of the position and momentum operators 
$\hat{X}_\mu$ and $\hat{P}_\mu$. The basic dynamical 
variables of our representation on 
the space of wavefunctions $\tilde\phi(p^\mu\vsi,x^\mu\vsi)$ 
are 
$\hat{X}^{\!\ssc \tilde{L}}= x+ i\partial_{p}= x\star$ and 
$\hat{P}^{\!\ssc \tilde{L}} = p - i\partial_{p}= p\star$,
where we have dropped the $\mu$ indices
and the subscript $\vsi$.  We may also write a general
function of $(p^\mu\vsi,x^\mu\vsi)$ as simply $\za(p,x)$, 
and  the $\star$ is as in the Moyal star product 
\bea
\za \star \zb (p,x) = \za(p,x) e^{-i (\cev{\partial}_p \vec{\partial}_x-\cev{\partial}_x \vec{\partial}_p) } \zb(p,x) \;,
\eea
with  $\za(p,x)\star=\za(p\star,x\star)$. The simplified 
notation is what we use in this and the next section without
any reference to the material from the $H_{\!\ssc R}(4)$
picture above. Under such notation, the story looks 
quite the same as the case for $H_{\!\ssc R}(3)$ with only  
$\hat{X}_i^{\!\ssc \tilde{L}}$, and  $\hat{P}_i^{\!\ssc \tilde{L}}$ 
as $x_i\star$ and $p_i\star$, given in details in Ref.\cite{070}.  
Hence, we present here only a summary of the results, leaving 
the readers to consult the latter paper and references therein. 

Let us take a little detour first to clarify our theoretical 
perspective. What we have is rather like the WWGM 
put up-side-down \cite{070}. We start with the quantum 
theory as an irreducible representation of a (quantum) 
relativity symmetry, including the Heisenberg-Weyl 
symmetry. With the wavefunction in the coherent 
state basis as the natural reduction of the representation 
of the group algebra, the corresponding representation 
of the latter properly extended serves as the algebra 
of observables. The latter can be seen as a collection 
of functions and tempered distribution of  the position 
and momentum operators represented as differential 
operators by $x\star$ and $p\star$. The real variables 
$x$ and $p$ are not quite the coordinates of the 
classical phase space. Only their rescaled counterparts 
under the contraction of the symmetry to the classical 
relativity symmetry are. Contrary to a deformation 
quantization, a contraction is a de-quantization 
procedure. From the algebraic point of view, the 
deformation of an observable algebra as in WWGM is 
really a result of a deformation of the classical 
relativity symmetry to the quantum one, pushed onto 
the group C$^*$-algebra of the symmetry. The contraction
is exactly the inverse of the deformation \cite{060},
at a Lie algebra level.
On the Hilbert space ${\mathcal{K}}$ of wavefunctions 
$\phi(p,x)$,  symmetries are represented in a form of 
unitary and antiunitary operators, factored by its closed 
center of phase transformations. On the set ${\mathcal{P}}$ 
of pure state density operators $\rho_\phi(p,x)\star$, 
corresponding to the abstract projection operator
$\hat\rho_\phi = \left|\phi\rra\!\lla \phi \right|$ for 
normalized $\left|\phi\rra$, the automorphism group 
$Aut({\mathcal{P}})$  is characterized by the subgroup 
of the group of real unitary transformations 
${\mathcal{O}}(\tilde{\mathcal{K}}_{\!\ssc R})$
compatible with the star product,  
$\tilde{\mathcal{K}}_{\!\ssc R}$ being the real 
span of  all $\rho_\phi(p,x)\star$, the complex 
extension of which is the Hilbert space of
Hilbert-Schmidt operators, as in the Tomita 
representation. We write the unitary 
transformations in the form
\[
 \tilde{U}_\star\za\star=\mu(\za)\star=U_\star\! \star \za  \star  \bar{U}_\star\!\star \;,
\]
with $\mu \in Aut({\mathcal{P}})$, where
${U}_\star\!\star \equiv {U}_\star(p,x)\star$
is a unitary operator on ${\mathcal{K}}$, generated 
by the Hermitian operator in the form of a real function 
$G_{\!s}(p\star,x\star)$, and $\bar{U}_\star\!\star$ is its 
inverse obtained by the complex conjugation and
$ \tilde{U}_\star \in {\mathcal{O}}(\tilde{\mathcal{K}}_{\!\ssc R})$. 
We refer to the ${U}_\star\!\star$ as star-unitary, in 
particular whenever necessary to highlight it being 
a function of the  $p\star$ and $x\star$ operators.

The above, illustrated for the case of  $H_{\!\ssc R}(3)$
formulation of standard quantum mechanics in Ref. \cite{070},
can be applied to our $H_{\!\ssc R}(1,3)$ case with a slight 
modification. We need to use the invariant inner product with 
$\hat\rho_\phi = \left.\left|\phi\rra\!\rra\!\lla\!\lla \phi \right|\right.$
for normalized $\left.\left|\phi\rra\!\rra$, and
replace the Hermitian and unitary requirements by
${\mathcal{P}}_4$-Hermitian and ${\mathcal{P}}_4$-unitary
ones.  Our relevant symmetry transformations are 
to be given by ${\mathcal{P}}_4$-unitary operator 
$V_{\star(s)} \star$ generated by
${\mathcal{P}}_4$-Hermitian $G_{\!s}(p\star,x\star)$,
which are real functions of  the 
basic ${\mathcal{P}}_4$-Hermitian  operators $(p\star,x\star)$,
{\em i.e.} $G_{\!s}(\hat{P}^{\!\ssc L}_{\!\ssc \mu},\hat{X}^{\!\ssc L}_{\!\ssc \mu})
   = \overline{G}_{\!s}(\hat{P}^{\!\ssc L}_{\!\ssc \mu},\hat{X}^{\!\ssc L}_{\!\ssc \mu})$,
and we use the $\bar\za$ to denote the `complex conjugate'
of  $\za$ as a function which correspond to $\bar\za\star$
as the ${\mathcal{P}}_4$-Hermitian conjugate of $\za\star$
as an operators as an element of observable algebra. The 
conjugation is the involution of the latter as a $^*$-algebra.
$\overline{V}_{\star(s)}\star$ of a ${\mathcal{P}}_4$-unitary 
${V}_{\star(s)}\star$ is to be interpreted in the same manner. 
The feature of $\overline{V}_{\star(s)}\star$ to be the inverse 
of ${V}_{\star(s)}\star$ is exactly a ${\mathcal{P}}_4$-unitarity.

Generators of our relativity symmetry $H_{\!\ssc R}(1,3)$ 
are to be represented as a subgroup of $Aut({\mathcal{P}})$ 
of the observable algebra. Formally, all expressions look
the same as if the ${\mathcal{P}}_4$-Hermitian and
${\mathcal{P}}_4$-unitary nature is not different from
the usual  Hermitian and unitary case. Again, the 
pseudo-unitary nature of the inner product does not
quite reveal itself in the essential coherent state 
representation. All $H_{\!\ssc R}(1,3)$ generators are 
${\mathcal{P}}_4$-Hermitian, hence each is given by a real
 $G_{\!s}$, generating (star-)${\mathcal{P}}_4$-unitary
${V}_{\star(s)}\!\star=e^{\frac{-i s}{2}{G}_{\!s}\star}$ 
as one-parameter groups of symmetry transformations.
Note that the factor $2$ is really $\hbar$. 
We have $\tilde{V}_{\star(s)}= e^{\frac{-i s}{2}\tilde{G}_{\!s}}$, 
\bea\label{vg}
 \tilde{V}_\star\za\star=\mu(\za)\star=V_\star\! \star \za  \star  \overline{V}_\star\!\star 
\eea
with 
\bea \label{tildeG}
\tilde{G}_{\!s} \rho 
    ={G}_{\!s}\!\star \rho -\rho \star\!{G}_{\!s} 
  = 2i \{{G}_{\!s},  \rho\}_\star\;,
\eea
where $\rho(p,x)\in  \tilde{\mathcal{K}}$ and 
$\{\cdot, \cdot\}_\star$  is the Moyal bracket. 
Hence, with $\rho(s)=\tilde{V}_{\star(s) }\rho(s=0)$, 
\bea \label{leq}
\frac{d}{ds}\rho(s)= \{{G}_{\!s},  \rho(s)\}_\star \;.
\eea
The equation  is the Liouville equation of motion for a 
mixed state $\rho$ in $\tilde{\mathcal{D}}$, the self-dual 
cone of $\tilde{\mathcal{K}}$. The class of operators on 
$\tilde{\mathcal{K}}$ representing symmetry generators 
are important, especially for tracing the symmetries to
 the classical limit where all ${G}_{\!s}\star$ reduce 
essentially to the commutative ${G}_{\!s}$, as multiplicative 
operators on the functional space of classical observables.
We can write $\tilde{G}_{\!s}=\hat{G}_{\!s}^{\ssc \tilde{L}}-\hat{G}_{\!s}^{\ssc \tilde{R}}$, 
where $\hat{G}_{\!s}^{\ssc \tilde{L}}\equiv {G}_{\!s}(p,x)\star
      ={G}_{\!s}(\hat{P}^{\!\ssc \tilde{L}},\hat{X}^{\!\ssc \tilde{L}})$ 
is a left action and $\hat{G}_{\!s}^{\ssc \tilde{R}}$
 is the corresponding right action defined by  
$\hat{G}_{\!s}^{\ssc  \tilde{R}}\za\equiv \za\star\!{G}_{\!s}(p,x) 
    ={G}_{\!s}(\hat{P}^{\!\ssc  \tilde{R}},\hat{X}^{\!\ssc  \tilde{R}})\za$.
Analogously to $\hat{X}^{\!\ssc  \tilde{L}}$ and $\hat{P}^{\!\ssc  \tilde{L}}$ 
coming from the left-invariant vector fields of the Heisenberg-Weyl 
group, there are those from the right-invariant ones given by
\bea
\hat{X}^{\!\ssc  \tilde{R}}= x -  i \partial_{p}\;,
\qquad
\hat{P}^{\!\ssc  \tilde{R}} = p +  i \partial_{x}\;.
\label{R}
\eea

From Eq.(\ref{v-shift}) we see that
\bea \label{phitrans}
V_{\star(-x'^\mu)}\!\star \tilde\phi (p^\mu,x^\mu)& =& e^{\frac{-i x'^\mu}{2} (-p_\mu\star)} \tilde\phi (p^\mu,x^\mu)
  = \tilde\phi \!\left(p^\mu,x^\mu+\frac{x'^\mu}{2} \right) e^{\frac{ix'_\mu p^\mu}{2}}\;,
\nonumber \\
V_{\star(p'^\mu})\!\star \tilde\phi (p^\mu,x^\mu)& =& e^{\frac{-i p'^\mu}{2}(x_\mu\star)} \tilde\phi (p^\mu,x^\mu) 
  =  \tilde\phi \!\left(p^\mu+\frac{p'^\mu}{2},x^\mu \right) e^{\frac{-ip'_\mu x^\mu}{2}}\;,
\eea
In the above,  for the wavefunctions, we show only the 
involved pair of variables in each case, and there is always 
no summation over indices. The other variables are simply
not affected by the transformations. In terms of the 
parameters  $x^\mu$ and $p^\mu$, we have
\bea
G_{\!-x^\mu}\star &=& p_\mu\star\;, 
\qquad\qquad 
\tilde{G}_{\!-x^\mu} =-2i \partial_{x^\mu} \;, 
\nonumber \\
G_{\!p^\mu}\star &=& x_\mu\star \;, 
\qquad\qquad   
\tilde{G}_{\!p^\mu} = 2i \partial_{p^\mu} \;,
\eea
all in the same form as in the $H_{\!\ssc R}(3)$ case.
The factors of $2$ in the translations $V_{\star(x)}\!\star$ 
and $V_{\star(p)}\!\star$, though somewhat suspicious 
at the first sight, are related to the fact that the arguments 
of the wavefunction correspond to half of the expectation 
values, due to our coherent state labeling. Thus, $x_\mu\star$ 
and $p_\mu\star$ generate translations of the expectation 
values, which is certainly the right feature to have.
For the Lorentz transformations, we have 
$G_{\!\omega^{\mu\nu}}=(x_\mu p_\nu-x_\nu p_\mu)$,
\bea
G_{\!\omega^{\mu\nu}}\star &=& (x_\mu p_\nu -i x_\mu \partial_{x^\nu} + i p_\nu \partial_{p^\mu}
    + \partial_{x^\nu} \partial_{p^\mu}) - (\mu \leftrightarrow \nu)\;,
\nonumber \\
\tilde{G}_{\!\omega^{\mu\nu}}  &=& -2i (x_\mu \partial_{x^\nu}- p_\nu \partial_{p^\mu} )- (\mu \leftrightarrow \nu)\;.
\eea
with the explicit action (no summation over the indices)
\begin{equation}
V_{\star(\omega^{\mu\nu})}\!\star \tilde\phi (p,x)
 = e^{\frac{-i\omega^{\mu\nu}}{2} (G_{\!\omega^{\mu\nu}}\star)} \tilde\phi (p,x)
    = \tilde\phi \!\left(\! e^{\frac{i\omega^{\mu\nu}}{2}  \widehat{G}_{\!\omega^{\mu\nu}} }[p,x] \!\right) ,
\end{equation}
where $\widehat{G}_{\!\omega^{\mu\nu}}$ are the  
infinitesimal $SO(1,3)$ transformation operators 
corresponding to the coset space action to be obtained
from Eq.(\ref{group}). The results are again in the 
same form as those for the $H_{\!\ssc R}(3)$ case.

All the  $G_{\!-x^\mu},\,G_{\!p^\mu}$ and $G_{\!\omega^{\mu\nu}}$ 
(and $G_{\!\theta}=1$) make the full set of operators 
for the generators  $\hat{G}_{\!s}^{\ssc L}=G_{\!s}\star$
of the $H_{\!\ssc R}(1,3)$ group representing the
symmetry on ${\mathcal{K}}$, and constitute a 
Lie algebra within the algebra of physical observables.
$\hat{G}_{\!s}^{\ssc R}$ set does the same as a right 
action, and $\hat{G}_{\!s}^{\ssc L}$ always commute 
with  $\hat{G}_{\!s'}^{\ssc R}$ since, in general,
$[ \hat\za^{\!\ssc \tilde{L}}, \hat\zc^{\!\ssc \tilde{R}}]=0$. 
These fourteen $G_{s}$ as multiplicative operators, 
of course, all commute among themselves. The  
commutators for $\tilde{G}_{\!s}$ are same as those 
for $\hat{G}_{\!s}^{\ssc L}$, with however the 
vanishing $\tilde{G}_{\!\theta}$  giving a vanishing 
$[\tilde{G}_{\!p^\mu}, \tilde{G}_{\!-x^\nu}]$.  
For any function $\za(p^\mu,x^\mu)$, there are four 
associated operators on $\tilde{\mathcal{K}}$. Those 
are $\za, \hat{\za}^{\!\ssc \tilde{L}},\hat{\za}^{\!\ssc \tilde{R}}$ 
and $\tilde{\za}$, but only two of them are linearly 
independent. For our relativity symmetry operators, 
the independent set $\{G_{\!-x^\mu},G_{\!p^\mu},G_{\!\omega^{\mu\nu}},
   \tilde{G}_{\!-x^\mu},\tilde{G}_{\!p^\mu},\tilde{G}_{\omega^{\mu\nu}}\}$
has the only non-vanishing commutators among 
them given by (we also have ${G}_{\!\theta}=1$, 
the identity, and  $\tilde{G}_{\!\theta}=0$)
\bea &&
[{G}_{\!\omega^{\mu\nu}}, \tilde{G}_{\!\omega^{\za \zb}}]
   = 2i( \eta_{\nu \zb} G_{\!\omega^{\mu \za }} -  \eta_{\nu \za } G_{\!\omega^{\mu \zb}}
                   +  \eta_{\mu \za } G_{\!\omega^{\nu  \zb}}-   \eta_{\mu \zb} G_{\!\omega^{\nu  \za }} ) \;,
\nonumber \\ &&
[{G}_{\!\omega^{\mu\nu}}, \tilde{G}_{\!-x^\za}] 
  =  -2i( \eta_{\nu\za} G_{\!-x^\mu} - \eta_{\mu \za} G_{\!-x^\nu} ) \;,
\nonumber \\ &&
[{G}_{\!\omega^{\mu \nu}}, \tilde{G}_{\!p^\za}]
  = -2i( \eta_{\nu \za} G_{\!p^\mu} - \eta_{\mu  \za} G_{\!p^\nu} ) \;,
\nonumber \\ &&
[\tilde{G}_{\!\omega^{\mu\nu}}, {G}_{\!-x^\za}]
   = -2i( \eta_{\nu\za} G_{\!-x^\mu} - \eta_{\mu \za} G_{\!-x^\nu} )  \;,
\nonumber \\ &&
[\tilde{G}_{\!\omega^{\mu\nu}}, {G}_{\!p^\za}]
   = -2i( \eta_{\nu \za} G_{\!p^{\mu}} - \eta_{\mu\za} G_{\!p^\nu} ) \;,
\nonumber \\ &&
[{G}_{\!p^\mu}, \tilde{G}_{\!-x^\nu}] =- [{G}_{\!-x^\mu}, \tilde{G}_{\!p^\nu}]
   = 2i \eta_{\mu\nu} \;,
\nonumber \\ &&
[{G}_{\!p^\mu}, \tilde{G}_{\!p^\nu}] =
[{G}_{\!-x^\mu}, \tilde{G}_{\!-x^\nu}] = 0\;.
\label{GtG}
\eea

Quantum dynamics is completely symplectic, whether
described in the Schr\"odinger picture in terms of
real/complex coordinates of the (projective) Hilbert
space or the Heisenberg picture which can be seen as
noncommutative coordinates description of the same
phase space \cite{078}. The explicit dynamical equation
of motion is to be seen as the transformations generated
by a physical Hamiltonian characterized by an evolution
parameter. In the $H_{\!\ssc R}(3)$ case of the usual 
(`non-relativistic') quantum mechanics, it is  
${G}_{\!t }= \frac{p_ip^i}{2m} +v(x^i)$.
For our  $H_{\!\ssc R}(1,3)$ case, we consider a
${G}_{\!\tau}=\frac{p_\mu p^\mu}{2m} +v(x^\mu)$
with the parameter 
$\tau$ being the Einstein proper 
time, which is expected to give Einstein particle 
dynamics  in the `free particle' case of vanishing
potential $v(x^\mu)$, as we  see explicitly below.

For some $s$-dependent operator $\za(p^\mu(s),x^\mu(s)) \star$ 
and a general Hamiltonian $G_{\!s}$, Heisenberg 
equation of motion is given by
\begin{equation}
    \frac{d}{ds}\za \star =\frac{1}{2i} \left[\za \star,G_{\!s}\!\star \right] \;.
\end{equation}
The right-hand side of the equation is
simply the Poisson bracket of  
$\za(p\star,x\star)$ and $G_{\!s}(p\star,x\star)$, 
functions of the noncommutative canonical variables
$p^\mu\star$ and $x^\mu\star$. The equation
can simply be written as 
\bea 
\frac{d}{ds}\za=\{\za,{G}_{\!s} \}_\star=\frac{-1}{2i}\tilde{G}_{\!s}\za \;,
\eea
and is exactly the differential version of the automorphism
flow given in Eq.(\ref{vg}), here with our ${\mathcal{P}}_4$-unitary
symmetry flows ${V}_{\star(s)}= e^{\frac{-i s}{2}{G}_{\!s}\star}$
generated by a  ${\mathcal{P}}_4$-Hermitian ${G}_{\!s}\star$. 
$\frac{-1}{2i}\tilde{G}_{\!s}$ is really a  Hamiltonian vector 
field for a Hamiltonian function $G_{\!s}(p\star,x\star)$ \cite{078}.

Our physical Hamiltonian operator ${G}_{\!\tau}(p\star,x\star)$ 
is  such a ${\mathcal{P}}_{\!\ssc 4}$-Hermitian ${G}_{\!s}\star$
provided that $v(x^\mu\star)$ is a real function of the four
$x^\mu\star$ operators. The corresponding Heisenberg 
equation gives, in particular,
\bea &&
\frac{d}{d\tau}x_\mu\star = \frac{1}{2i} \frac{1}{2m}[x_\mu\star,p^\mu\star p_\mu\star]
=\frac{p_\mu\star}{m} =\frac{\partial G_{\!\tau}(p\star,x\star)}{\partial (p^\mu\star)} \;,
\sea
\frac{d}{d\tau}p_\mu\star = \frac{1}{2i}[p_i\star,v(x^\mu\star)]
= -\frac{\partial v(x^\mu\star) }{\partial (x^i\star)} 
=-\frac{\partial G_{\!\tau}(p\star,x\star)}{\partial (x^\mu\star)} \;.
\eea
which are exactly
\bea\label{heq}
\frac{d}{d\tau}  \hat{X}_\mu^{\!\ssc \tilde{L}} 
=\frac{\partial G_{\!\tau}(\hat{P}_\nu^{\!\ssc \tilde{L}},\hat{X}_\nu^{\!\ssc \tilde{L}}  )}{\partial \hat{X}^{{\!\ssc \tilde{L}}^\mu} } \;,
\qquad
\frac{d}{d\tau}  \hat{P}_\mu^{\!\ssc \tilde{L}} 
=-\frac{\partial G_{\!\tau}(\hat{P}_\nu^{\!\ssc \tilde{L}},\hat{X}_\nu^{\!\ssc \tilde{L}}  )}{\partial \hat{P}^{{\!\ssc \tilde{L}}^\mu} } \;,
\eea
in the standard form of Hamilton's equations of motion
for the canonical ${\mathcal{P}}_{\!\ssc 4}$-Hermitian
operator coordinate pairs 
$\hat{X}_\mu^{\!\ssc \tilde{L}}$-$\hat{P}_\mu^{\!\ssc \tilde{L}}$.
 With the vanishing potential,  we have 
 $G_{\!\tau}\star=\frac{1}{2m} [-(\hat{P}_0^{\!\ssc \tilde{L}})^2 + \sum_i (\hat{P}_i^{\!\ssc \tilde{L}})^2]$
resulting in $\hat{P}_\mu^{\!\ssc \tilde{L}}$ 
being $\tau$ independent and 
 $\frac{d\hat{X}_\mu^{\!\ssc \tilde{L}}}{d\tau}
      = \frac{1}{m}\hat{P}_\mu^{\!\ssc \tilde{L}}$,
which is the Einstein relation of four-momentum being equal 
to the Einstein four-velocity multiplied by the particle mass.

For the Schr\"odinger picture, as ${\mathcal{P}}_{\!\ssc 4}$-unitary
flows on $\mathcal{K}$, we have the equation
\begin{equation}
    \frac{d}{ds}\tilde\phi=\frac{1}{2i}G_{\!s}\star \tilde\phi \;,
\end{equation}
which, for $G_{\!\tau}\star$ with a vanishing potential, 
is in the exact form of the Klein-Gordon equation and 
gives the $\tau$-independent solution for $\tilde\phi$,
provided that the $G_{\!\tau}\star$ eigenvalue is taken 
to be $-\frac{mc^2}{2}$.  Explicitly, in terms
of the basic variables $p^\mu$ and $x^\mu$, we have
\begin{equation}
    G_{\!\tau}\star \tilde\phi(p,x) 
=\frac{1}{2m} p_\mu\star p^\mu\star \tilde\phi(p,x) 
=\frac{1}{2m} \left(p^\mu p_\mu- \eta^{\mu\nu}\partial_{x^\mu}\partial_{x^\nu}
    -2ip^\mu\partial_{x^\mu}\right)\tilde\phi(p,x)
\;,
\end{equation}
giving the free-particle wavefunctions 
$\tilde\phi(p,x)=e^{i(2k_\mu-p_\mu) x^\mu}$ for eigenvalues 
$2k^\mu k_\mu$.  Eigenvalues of the momentum operators 
$p_\mu\star$ are $2k_\mu$, satisfying $(2k^\mu) (2k_\mu)=-m^2c^2$.  
The factor of 2 really corresponds to $\hbar$, as in the 
standard textbook expression. Finally, the $\tau$-dependence 
is then given by 
$\frac{d}{d\tau}\tilde\phi = -\frac{m c^2}{2i} \tilde\phi$,
as expected.

We have studied in details the covariant harmonic oscillator 
problem under the same ${\mathcal{P}}_{\!\ssc 4}$-unitarity,
though on the wavefunction formulated in the `position
eigenstate' basis,\cite{083}. Some of the corresponding 
results under the wavefunction in coherent state basis, 
$\tilde\phi(p^\mu,x^\mu)$, can be found in the appendix.
Solution to the problem, as well as for the Einstein
particle described above, shows a successful 
applications and hence the validity of the 
theoretical construction.

\subsection{Lorentz to Galilean Contraction} \label{LtoG2}
Contraction to Galilean limit has been presented in 
Sec. \ref{sec4} for algebra of operators on the Hilbert 
space and the group representation in a form of wavefunctions 
in the coherent state basis. In this section, we present 
the corresponding contraction in the observable algebra 
given in the WWGM formalism, described above. Recall that 
the original Hilbert space under the contraction becomes 
reducible into a sum of essentially identical irreducible 
components, each being spanned by the wavefunctions 
$\tilde\phi(p^i,x^i) \equiv \tilde\phi (p\shi^i, x\shi^i)$
for a particular value of `time' ${t}_{\!\ssc (\!\chi\!)}$.
A general operator 
$\za(\hat{X}_{\mu}^{\!\ssc \tilde{L}},\hat{P}_{\mu}^{\!\ssc \tilde{L}})$
should then be seen as 
$\za(\hat{X}_{i}^{\!\ssc \tilde{L}},\hat{P}_{i}^{\!\ssc \tilde{L}},
   \hat{T}^{\ssc\! \tilde{L}}_{\!\ssc (\!\chi\!)},
   \hat{H}^{\ssc\! \tilde{L}}_{\!\ssc (\!\chi\!)})$ with 
$\hat{X}_{i}^{\!\ssc \tilde{L}} \equiv \hat{X}^{\ssc\! \tilde{L}}_{\!\ssc (\!\chi\!)i}$
and $\hat{P}_{i}^{\!\ssc \tilde{L}} \equiv \hat{P}^{\ssc\! \tilde{L}}_{\!\ssc (\!\chi\!)i}$,
from results of Eq.(\ref{XPTH}). Hence, on $\tilde\phi(p^i,x^i)$ 
we have effectively Hermitian actions of operators
$\hat{X}_{i}^{\!\ssc \tilde{L}} = x_i + i \partial_{p^i}$,
$\hat{P}_{i}^{\!\ssc \tilde{L}} = p_i - i \partial_{x^i}$,
$\hat{T}^{\ssc\! \tilde{L}}_{\!\ssc (\!\chi\!)} \to {t}_{\!\ssc (\!\chi\!)}$,
and $\hat{H}^{\ssc\! \tilde{L}}_{\!\ssc (\!\chi\!)} \to {e}_{\!\ssc (\!\chi\!)}$,
with the last two reduced to a simple multiplication 
by the `variables'  ${t}_{\!\ssc (\!\chi\!)}$ and 
(formally infinite) ${e}_{\!\ssc (\!\chi\!)}$, 
respectively. All $\za(p^\mu\star,x^\mu\star)$ 
operators on $\tilde\phi(p^i,x^i)$ reduce to 
$\za(p^i\star,x^i\star, {t}\shi, {e}\shi)$, or rather 
simply to $\za(p^i\star,x^i\star)$ like in the basic 
quantum mechanics, a unitary representation theory of 
$H_{\!\ssc R}(3)$.  The $\star$ should now be seen as 
the one  involving only variables $p^i$ and $x^i$. 

The transformations generated by the Hermitian
$G_{\!-x^i}\star,G_{\!p^i}\star$ and $G_{\!\omega^{ij}}\star$ 
obviously do not change. They represent generators of 
the $H_{\!\ssc R}(3)$ subgroup of $H_{\!\ssc R}(1,3)$ 
to begin with. $\tilde{G}_{\!-x^i},\tilde{G}_{\!p^i}$ 
and $\tilde{G}_{\!\omega^{ij}}$ are also unchanged. 
$G_{\!-x^{\ssc 0}}\star$ and $G_{\!p^{\ssc 0}}\star$, 
representing $\hat{P}_{\ssc (\!\varsigma\!)0}^{\!\ssc \tilde{L}}$
and $\hat{X}_{\ssc (\!\varsigma\!)0}^{\!\ssc \tilde{L}}$,
are to be replaced under the contraction by
$\hat{H}^{\ssc\! \tilde{L}}_{\!\ssc (\!\chi\!)}$ and
$\hat{T}^{\ssc\! \tilde{L}}_{\!\ssc (\!\chi\!)}$, respectively, with
 $V_{\star(-x^{\ssc 0})} =e^{\frac{i x^{\ssc 0}}{2}G_{\!-x^{\ssc 0}}}$
and $V_{\star(p^{\ssc 0})} =e^{\frac{-i p^{\ssc 0}}{2}G_{\!p^{\ssc 0}}}$ 
re-expressed as $V_{\star(t)} =e^{-\frac{it}{2}G_{\!t}}$
and $V_{\star(e)} =e^{\frac{ie}{2}G_{\!-e}}$, 
where $G_{\!t}\star=\hat{H}^{\ssc\! \tilde{L}}_{\!\ssc (\!\chi\!)}$
and $G_{\!-e}\star=\hat{T}^{\ssc\! \tilde{L}}_{\!\ssc (\!\chi\!)}$. 
On the wavefunction $\tilde\phi(p^i,x^i)$, we have the infinite
$G_{\!t}\star={e}\shi$ and finite $G_{\!-e}\star={t}\shi$. 
We also have $\tilde{G}_{\!t}= 2i \partial_{{t}\shi}$ and
$\tilde{G}_{\!-e} = -2  i\partial_{{e}\shi}$. 
None of the four operators are of interest, so long 
as their action on the observable algebra for an 
irreducible representation $\tilde\phi(p,x)$ is concerned.

The other interesting ones to check are the Lorentz boosts 
under the contraction. The generator $J_{{\ssc 0}i}$ 
in the Lie algebra is replaced by the finite $K_i=
\frac{1}{c}J_{{\ssc 0}i}$. The group elements
$e^{i{\omega^{0i}}J_{{\ssc 0}i}}$ are to be re-expressed as
$e^{i{\zb^i} K_i}$ with ${\zb^i}= c \, {\omega^{0i}}$. 
In the original representation, the $J_{{\ssc 0}i}$ action is given by 
$G_{\!\omega^{0i}}\star= \hat{X}_{\ssc (\!\varsigma\!)0}^{\!\ssc \tilde{L}} 
  \hat{P}_{\ssc (\!\varsigma\!)i}^{\!\ssc \tilde{L}}  - \hat{X}_{\ssc (\!\varsigma\!)i}^{\!\ssc \tilde{L}}  
  \hat{P}_{\ssc (\!\varsigma\!)0}^{\!\ssc \tilde{L}}$, 
from which follows the action of $K_i$ as
\[
G_{\!\zb^i} =  -\hat{T}^{\ssc\! \tilde{L}}_{\!\ssc (\!\chi\!)} \hat{P}_{i}^{\!\ssc \tilde{L}} 
     - \hat{X}_{i}^{\!\ssc \tilde{L}} \left(\frac{-1}{c^2}\hat{H}^{\ssc\! \tilde{L}}_{\!\ssc (\!\chi\!)} \right)
   \to - {t}_{\!\ssc (\!\chi\!)} p_i\star 
   = {t}_{\!\ssc (\!\chi\!)} G_{\!-x^i}
\]
with $V_{\star(\zb^i)}= e^{\frac{-i \zb^i}{2} G_{\!\zb^i}}$
(no summation over $i$), a re-writing of
$V_{\star(\omega^{0i})}$ with the new finite parameter 
$\zb^i$. We have seen, in Eq.(\ref{phitrans}) explicitly,  
that $V_{\star(-x^i)}\star$ gives a translation in the 
variable $x^i$ of the wavefunction. $V_{\star(\zb^i)}\star$ 
is then a time variable ${t}_{\!\ssc (\!\chi\!)}$-dependent 
translation, a Galilean boost exactly as the Lie 
algebra contraction promised, and is now unitary.
Similarly, we have
\bea
\tilde{G}_{\!\zb^i} = \frac{1}{c}\tilde{G}_{\!\omega^{0i}}
 &=& -2i (-t\shi \partial_{x^i} - p_i \partial_{e\shi})  +\frac{2i}{c^2} (x_i\partial_{t\shi} +e \partial_{p^i})
\sea
\rightarrow 2i (t\shi \partial_{x^i} +p_i \partial_{e\shi}) \;.
\eea
We keep the $\partial_{e\shi}$ since the $\tilde{G}_{\!\zb^i}$ 
may act on the mixed states. We have the newly relevant nonzero
commutators involving a ${G}_{\!\zb^i}$, ${G}_{\!t}$, or
$G_{\!-e}$, and a $\tilde{G}_{\!s}$ as well as those 
involving a $\tilde{G}_{\!\zb^i}$, $\tilde{G}_{\!t}$, or 
$\tilde{G}_{\!-e}$ and a ${G}_{\!s}$, all from the
generators of the Lie algebra, as
\bea &&
[{G}_{\!\zb^i}, \tilde{G}_{\!\omega^{\ssc jk}}]
=-2i\left(  \delta_{\ssc ij} G_{\!\zb^k}-\delta_{\ssc ik} G_{\!\zb^j}\right)\; ,
\sea  
 [{G}_{\!\omega^{\ssc ij}}, \tilde{G}_{\!\zb^k}]
=2i\left(  \delta_{\ssc ik} G_{\!\zb^j}-\delta_{\ssc jk} G_{\!\zb^i}\right)\;,  
\sea
[{G}_{\!\zb^i}, \tilde{G}_{\!t}]=[\tilde{G}_{\!\zb^i}, G_{\!t}]=2iG_{\!-x^{\ssc i}} \; , 
\sea
[{G}_{\!\zb^i}, \tilde{G}_{\!p^{\ssc j}}]=[\tilde{G}_{\!\zb^i},{G}_{\!p^{\ssc j}}]=2i\delta_{\ssc ij}G_{\!-e} \;,
\sea
[{G}_{\!-e}, \tilde{G}_{\!t}] 
=- [{G}_{\!t}, \tilde{G}_{\!-e}]=-2i \;.
\eea

Since on the Hilbert space of the contracted theory we have
only $\tilde\phi(p^i,x^i)$ and the corresponding observable 
algebra as $\za(p_i\star,x_i\star)$, the loss of 
$p_{\ssc 0}\star$ and $x_{\ssc 0}\star$, the quantum 
observables of energy and time, means that the Heisenberg 
equation of motion, in the form of a differential equation in $\tau$, 
 effectively corresponds to the part of $G_{\!\tau}\star$
involving only $p^i\star$ and $x^i\star$. We have
\bea
\frac{d}{d\tau} \za\star = \frac{1}{2i} [\za\star, G_{\!\tau}\star]
= \frac{1}{2i} [\za\star, G_{t\shi}\star] 
\eea
where $G_{t\shi}= \frac{p^i p_i}{2} + v(x^i)$, giving the
right time evolution in the `non-relativistic', or $H{\!\ssc R}(3)$,
quantum theory, as expected. At the $c \to \infty$ limit, the 
proper time is just the Newtonian time. One can also see that
the quantum Poisson bracket $\frac{1}{2i}[\cdots,\cdots]$ 
does suggest that the now multiplicative operators $t\shi$ 
and $e\shi$, from the original  $p_{\ssc 0}\star$ and 
$x_{\ssc 0}\star$, are to be dropped from the  canonical 
coordinates of the noncommutative symplectic geometry, 
in line with the Hilbert space picture.

\section{Contraction to Classical Theory in Brief} \label{sec6}

In this section we look at the corresponding classical theory 
at the Lorentz covariant level through the contraction 
along the line of the one performed in the `non-relavitistic', 
$H_{\!\ssc R}(3)$, case presented in Ref.\cite{070}. Only a 
sketch will be presented where the mathematics is essentially 
the same with the latter. The contraction trivializing the 
commutators between the position and momentum operators 
is obtained by rescaling the generators as 
\bea
X_\mu^c = \frac{1}{k_x} X_\mu
\qquad \mbox{and} \qquad
\label{XPc}
P_\mu^c =  \frac{1}{k_p} P_\mu \;,
\eea
and taking the limit $k_x, k_p \to \infty$. The only important 
difference between $k_x$ and $k_p$ parameters is their physical 
dimensions, giving the $X_\mu^c$ and $P_\mu^c$ observables 
with their different classical units.
For the corresponding operators we have
\bea
\hat{X}^{c\ssc L} &=&    x^c +   i \frac{1}{k_x k_p} \partial_{p^c} \longrightarrow x^c\;,
\nonumber \\
\hat{P}^{c\ssc L} &=&   p^c -   i \frac{1}{k_x k_p} \partial_{x^c}\longrightarrow p^c\;,
\eea
and the Moyal star-product reduces to a simple commutative product. 
Functions $\za(p\star,x\star)$, representing quantum observables,
reduce to multiplicative operators  $\za(p^c,x^c)$, the classical 
observables acting on the contracted representation space of the 
original pure and mixed states.

For the Hilbert space of pure states, the coherent
state basis is taken with the new labels as $\left|{p}^c,{x}^c\rra$, 
where  $2{p}_\mu^c$ and $2{x}_\mu^c$ characterize the 
expectation values of $\hat{X}_\mu^c$ and $\hat{P}_\mu^c$ 
operators. We have
\bea
\lla\!\!\lla {p}'^c_\mu,{x}'^c_\mu | \hat{X}_\mu^c |{p}_\mu^c,{x}_\mu^c\rra\!\!\rra
&=& [({x}'^c_\mu+{x}^c_\mu)-i({p}'^c_\mu-{p}^c_\mu)]
\lla\!\!\lla {p}'^c_\mu,{x}'^c_\mu \right.  \left|{p}_\mu^c,{x}_\mu^c\rra\!\!\rra \;,
\nonumber \\
\lla\!\!\lla {p}'^c_\mu,{x}'^c_\mu | \hat{P}_\mu^c |{p}_\mu^c,{x}_\mu^c\rra\!\!\rra
&=& [({p}'^c_\mu+{p}^c_\mu)+i({x}'^c_\mu-{x}^c_\mu)]
\lla\!\!\lla {p}'^c_\mu,{x}'^c_\mu \right.  \left|{p}_\mu^c,{x}_\mu^c\rra\!\!\rra \;,
\eea
with $\lla\!\!\lla {p}'^c_\mu,{x}'^c_\mu \right.  \left|{p}_\mu^c,{x}_\mu^c\rra\!\!\rra$  
at the contraction limit
going to zero for two distinct states. The Hilbert space, 
as a representation for the contracted symmetry, as well as
a representation of the now commutative algebra of observables, 
reduces to a direct sum of one-dimensional representations 
of the ray spaces of each $\left|{p}_\mu^c,{x}_\mu^c\rra$. 
The only admissible pure states are the exact coherent states, 
and not any linear combinations.  The obtained coherent states 
can be identified as  classical states, on the space of which 
the $\tilde{G}_{\!s}$-type operators act as generators of 
symmetries.  ${G}_{\!s}\star$-type operators, as  general 
$\za\star$ in the original observable algebra, contract to 
commuting multiplicative operators corresponding to classical 
observables. Results suggest that the projective Hilbert space, 
the true quantum phase space, in classical limit gives exactly 
the classic phase space with ${p}_\mu^c$ and ${x}_\mu^c$ 
coordinates. The Hilbert space, or Schr\"odinger picture 
otherwise, at the classical limit serves rather as the 
Koopman-von Neumann formulation in a broader setting of mixed 
state, {\em i.e.} statistical mechanics. We do not intend to 
explore that aspect further in this article. The observable 
algebra, or Heisenberg picture, gives a much more direct way 
of examining the full dynamical theory at that contraction 
limit. It also gives a direct and intuitive picture of the 
phase space geometry too. The original position and momentum 
operators, ${x}_\mu\star$ and ${p}_\mu\star$, can be seen as 
noncommutative coordinates of the noncommutative symplectic 
geometry which is nothing other than the projective Hilbert
space itself \cite{078}, described in a different way. The 
contracted versions as ${x}_\mu^c$ and ${p}_\mu^c$ are the 
classical phase space coordinates with no noncommutativity left.

Let us turn to the noncommutative Hamiltonian 
transformations. As mentioned above, at the quantum level, 
a ${G}_{\!s}\star={G}_{\!s}({p}_\mu\star, {x}_\mu\star)$
operator is a Hamiltonian function of the phase space 
coordinates $p\star$ and $x\star$, and the corresponding 
$\frac{-1}{2i}\tilde{G}_{\!s}$ is the Hamiltonian vector field.
It is, of course, well known since Dirac that what has now been
 identified as a quantum Poisson bracket $\frac{1}{2i}[\cdot,\cdot]$
 \cite{078,081} (and see references therein) reduces exactly to 
a classical Poisson bracket, which works in our formulation, 
explicitly shown in Ref.\cite{070}; {\em i.e.}
\[
 {G}_{\!s}({p}_\mu\star, {x}_\mu\star)  \to {G}_{\!s}^c({p}_\mu^c,{x}_\mu^c)\;,
\qquad
 \frac{-1}{2i}\tilde{G}_{\!s}=\frac{1}{2i}[\cdot,\cdot]
\to \{\cdot,{G}_{\!s}^c\}= \frac{-1}{2i}\tilde{G}_{\!s}^c \;.
\]
The explicit expressions are in exactly the same form as 
those of the quantum case, namely
\bea &&
\tilde{G}_{\!\omega^{\mu\nu}}^c= \tilde{G}_{\!\omega^{\mu\nu}}
=- 2i(x_\mu^c \partial_{x^{c\nu}} -p_\nu^c \partial_{p^{c\mu}}) 
    -(\mu \leftrightarrow \nu) \;,
\sea 
\tilde{G}_{\!-x^{c\mu}} =-2i \partial_{x^{c\mu}} \;, 
\qquad\qquad   
\tilde{G}_{\!p^{c\mu}} = 2i \partial_{p^{c\mu}} \;.
\eea
Note their independence on the contraction parameter $k$ (or $k_p$
and $k_x$), even before the $k \to \infty$ limit is 
explicitly taken. In conclusion, from the 
quantum Poisson bracket in terms of the Moyal bracket, or 
the Hamiltonian vector field given in terms of $\tilde{G}_{\!s}$,
we retrieve the Hamiltonian flow equation
\bea
\frac{d}{ds} \za(p^c,x^c) = \{ \za(p^c,x^c), {G}_{\!s}^c\} = \frac{-1}{2i}\tilde{G}_{\!s}^c \za(p^c,x^c) \;
\eea 
for any classical observable $\za(p^c,x^c)$ as a function 
of basic observables $x^{c\mu}$ and $p^{c\mu}$, which also
serve as canonical coordinates for the phase space, with
the standard expression for the classical Poisson bracket. 
The Hamilton's equations (\ref{heq}), as specific example,
become
\bea
\frac{d}{d\tau}x^c_\mu =\frac{\partial G_{\tau}^c}{\partial p^{c\mu}} = \frac{p^c_\mu}{m}
\qquad
\frac{d}{d\tau}p^c_\mu =-\frac{\partial G_{\tau}^c}{\partial x^{c\mu}}
  = -\frac{\partial v(x^{c\nu})}{\partial x^{c\mu}}\;.
\eea
$G_{\tau}^c=\frac{p^{c\mu}p^c_\mu}{2m} + v(x^{c\mu})$
is the covariant classical Hamiltonian, the $v=0$ case of
which is free particle dynamics in Einstein special relativity.

\section{Discussions and Conclusions}

We presented above the theory of symplectic dynamics 
as essentially an irreducible component of the regular 
representation of  the $H_{\!\ssc R}(1,3)$ (quantum) 
relativity symmetry, with a pseudo-unitary inner product 
obtained from that an earlier study of the covariant 
harmonic oscillator problem identified as a representation
of the same symmetry.  The explicit form of the inner 
product for the wavefunctions $\tilde\phi(p^\mu,x^\mu)$ 
is also given. Though the wavefunctions are divergent 
at timelike infinity, the inner product is always finite, 
without the need for domain restrictions. The quantum 
theory in terms of such wavefunctions is well behaved, 
and has no divergence for all physical quantities, again
without any artificial manipulation of the integrals 
which may not be mathematically sound. That is a success 
of the presented version of Lorentz covariant quantum 
mechanics that other formulations failed to achieve. 
Here, we are talking about a fully Lorentz covariant 
theory with the four-vector position and momentum 
operators $\hat{X}_\mu$ and $\hat{P}_\mu$.

Our study is a part of our quantum relativity 
group-theoretically based program. The constructed 
quantum mechanics is the `relativistic' version of 
the so-called `non-relativistic' theory based on the 
$H_{\!\ssc R}(3)$ group, or on the $\tilde{G}(3)$ group,  
a $U(1)$ central extension  of the Galileian group.  
$H_{\!\ssc R}(3)$ is a subgroup of the $H_{\!\ssc R}(1,3)$ 
group, while together with $\tilde{G}(3)$ they are both 
subgroup of the $c \to \infty$ approximation of the 
$H_{\!\ssc R}(1,3)$, obtained as a symmetry, or Lie 
algebra, contraction. Note that $H_{\!\ssc R}(3)$ 
is isomorphic to the $\tilde{G}(3)$ group with 
the one parameter subgroup of `time-translation' 
taken out. The $H_{\!\ssc R}(3)$ picture is an exact 
`time-independent' representation of $\tilde{G}(3)$ 
picture. Just like the earlier study of our group 
establishing the full Newtonian theory from the other 
contraction of the $H_{\!\ssc R}(3)$ quantum theory, 
we describe here how the `non-relativistic' quantum 
theory and  `relativistic' classical theory are to be 
retrieved successfully from the proper contractions.  

We skip the details of the reducible Tomita representation 
on the Hilbert space containing a self-dual cone of vectors 
corresponding to the mixed, statistical states, and leave 
it to the possible future analysis. For the Schr\"odinger 
evolution on the Hilbert space of pure states, the 
contraction of the quantum theory to its classical 
approximation is also easier to appreciate from the 
Tomita representation picture. Anyway, we have been very 
brief on the classical approximation analysis, since it 
is essentially an exact parallel of the earlier 
$H_{\!\ssc R}(3)$ case.

The focus of the paper is mostly on the formulational 
aspects. The key theme is our full group-theoretical
construction scheme from the identified relevant
relativity symmetries and the rigorous connections
between theories at different levels with the lower
ones as approximations to the higher ones along 
a symmetry contraction scheme. Our results here,
together with the earlier studies, give the successful 
implementation of all that for the `relativistic'  
quantum theory, down to its corresponding classical theory 
and the `non-relativistic' quantum and classical theories.  
Under that key theme, all four theories are the lower 
levels or approximations of the top level theory with 
a stable quantum relativity  \cite{030,071} symmetry, 
one that cannot be the contraction of another symmetry. 
The formulation of the current $H_{\!\ssc R}(1,3)$
theory, being essentially the only spin zero representation,
as a theory of a Lorentz covariant quantum mechanics with
the adoption of the pseudo-unitary inner product is of
special interest.

We believe the construction of this kind of fundamental 
theory presented here, based on a non-unitary representation,
has not been available in the literature. And the result
is considered a great success. Going along the logic of our
line of thinking \cite{082}, it is actually not a surprise at 
all. The bottom line is that going from the $H_{\!\ssc R}(3)$ 
theory to the $H_{\!\ssc R}(1,3)$ one is about going from 
the noncommutative geometry of the three-vectors 
$\hat{X}_i$ and $\hat{P}_i$ to that of the Minkowski 
four-vectors $\hat{X}_\mu$ and $\hat{P}_\mu$. The way the 
three-dimensional Euclidean space sits inside the four 
dimensional Minkowski spacetime as a pseudo-Euclidean 
space should be the basic guiding principle. And that 
can be seen just from the structure of the symmetry groups. 
The vector space spanned by the $\hat{X}_\mu$, for example, 
is a $1+3$ dimensional space of the Lorentz symmetry with 
the pseudo-Euclidean, therefore non-unitary, norm. The 
covariant harmonic oscillator system is a good place to 
start attacking the problem. In general, the harmonic 
oscillator problem is the most important prototype 
problem for any dynamical theory. For a quantum theory, 
the solution Fock states give one of the most useful 
basis for understanding the Hilbert space of pure states. 
The subspace spanned by the Fock states of a fixed $n$  
eigenvalue of the total number operators, under the 
right theory, corresponds exactly to the space of 
symmetric $n$-tensors. For the Lorentz symmetry, those 
are the symmetric products of the complexified basic 
pseudo-unitary Minkowski vector representation,
as formulated and analyzed in Ref.\cite{083} in the 
$x$-representation, namely wavefunctions $\phi(x^a)$. 
The anti-Hermitian operators $\hat{X}_{\!\ssc 0}$
and $\hat{P}_{\!\ssc 0}$  were represented as 
$i\hat{X}_{\!\ssc 4} \to ix_{\!\ssc 4}$ and 
$i\hat{P}_{\!\ssc 4} \to -i\hbar \partial_{x^{\ssc 4}}$,
respectively. We adapt that here to a $H(4)$ coherent 
state formulation, eventually obtaining the 
$\tilde\phi(x^\mu, p^\mu)$ representation of 
$H_{\!\ssc R}(1,3)$ symmetry.  The result is very
interesting, and in hindsight quite natural to expect.

The story is again easier to understand looking at the 
Fock state basis now given by the $\tilde\phi_n(x^\mu, p^\mu)$ 
wavefunctions, with $n$ being a shorthand notation for 
 $(n_{\ssc 0};n_{\ssc 1},n_{\ssc 2},n_{\ssc 3})$.
The four $n=1$ states are described by
$(x_\nu -ip_\nu) \tilde\phi_{\ssc 0}$, where 
$\tilde\phi_{\ssc 0}=e^{-\frac{x^\mu x_\mu +p^\mu p_\mu}{2}}$
is the Lorentz invariant $n=0$ wavefunction.  We can denote
those four wavefunctions as $\tilde\phi_{\ssc 1\nu}$ and  
identify them as components of a four-vector. For the norm
of the latter to be Minkowski, the  integrand of the inner 
product needs to contain
$\eta^{\rho\nu}\tilde\phi^*_{\ssc 1\rho} \tilde\phi_{\ssc 1\nu}$
which is, in particular, equal to $(x^\nu +ip^\nu)  (x_\nu -ip_\nu) |\tilde\phi_{\ssc 0}|^2$.
The $\tilde\psi^*(-x^{\ssc 0}, -p^{\ssc 0},x^i,p^i) \tilde\phi_n(x^\mu, p^\mu)$
product, instead of the usual 
$\tilde\psi^*(x^\mu, p^\mu) \tilde\phi_n(x^\mu, p^\mu)$,
calls for taking the $\hat{X}_{\!\ssc 0}$ and $\hat{P}_{\!\ssc 0}$, 
given by $x_{\ssc 0}+i\partial_{p^{\ssc 0}}$ and
$p_{\ssc 0}-i\partial_{x^{\ssc 0}}$
to be anti-Hermitian. The $\tilde\phi_{\ssc 0}$ factor is 
contained in all Fock state wavefunctions $\tilde\phi_n$, 
therefore also in all wavefunctions to be given by their 
finite linear combinations. However, unlike in $H_{\!\ssc R}(3)$ 
case, the zero-state wavefunction does not dictate the 
rapidly decreasing nature of the wavefunctions, which 
would be achieved by 
$e^{-(x^{\!\ssc 0})^2 -(p^{\!\ssc 0})^2}\tilde\phi_{\ssc 0}$. 
Instead of having the required extra factor included 
as a part of the wavefunction, putting it into the 
definition of the inner product keeps the right
Lorentz transformation properties of the $\tilde\phi_n$
wavefunctions. In hindsight, that can lead to the idea
of replacing the usual integral expression of the inner 
product by the one given in expression (\ref{ptip}).
The same reasoning can be applied to obtain the analogous 
pseudo-unitary inner product in $x$-representation with  
$\hat{X}_\mu$ and $\hat{P}_\mu$ as $x_\mu$ and 
$-i\hbar \partial_{x^\mu}$.

For wavefunction representation, we usually see the operators
of the form $x$ and $-i\hbar \partial_{x}$, or $x+i\partial_{p}$ 
and $p-i\partial_{x}$, for real variables $x$ and $p$ as Hermitian.  
But it should be understood that the naive notion of Hermiticity 
is defined with respect to the usual integral inner product.
That whole scheme does not work well for `coordinates' 
of a nontrivial metric signature for the case of which the
usually taken inner product does not respect the pseudo-orthogonal
rotational symmetry. Under our pseudo-unitary representation, 
the naively anti-Hermitian starting form of the 
$\hat{X}_{\!\ssc 0}$ and $\hat{P}_{\!\ssc 0}$ operators
is eventually realized in the naively Hermitian form.
Either way, they are truly Hermitian, together with the 
$\hat{X}_i$ and $\hat{P}_i$, meaning they are self-adjoint 
with respect to the inner product. We have also identified
that as the ${\mathcal{P}}_{\!\ssc 4}$-Hermiticity, in
line with the notion of pseudo-Hermitian quantum
mechanics. Again, that pseudo-Hermiticity is the true
Hermiticity with respect to the proper inner product. 
Only the corresponding pseudo-unitarity is definitely not 
a unitarity in the sense that the inner product does not
give a positive definite norm.  

We start by exploring the formulation without worrying
much about the positivity of eigenvalues and the feasibility 
of a probability picture.  As commented in Ref.\cite{082,083},
we see those as unimportant, with the availability of 
a new perspective on quantum physics in terms of clearly
defined symplectic dynamics and the noncommutative
values of the observables \cite{078,079,081}. The bottom
line, though, is that even if the standard probability 
interpretation for the usual quantum theory is required, 
its analog for a spacetime quantum theory is not at all 
justified. The notion of probability for finding a particle 
somewhere in space at a fixed time hardly has a spacetime 
analog, nor do we have any solid way to understand, not 
to say implement, von Neumann measurement in a spacetime
theory. However, the current study certainly shows that 
the $\hat{X}_{\!\ssc 0}$ and $\hat{P}_{\!\ssc 0}$ operators 
have the positive eigenvalues, as usually expected.

As mentioned above, the $\hat{X}_\mu$ and $\hat{P}_\mu$ 
operators can be taken as noncommutative coordinates of 
the quantum phase space, which is otherwise described 
in real/complex number coordinates as the infinite dimensional 
projective Hilbert space. That picture has been solidly
established in Ref. \cite{078,081} for the $\hat{X}_i$ 
and $\hat{P}_i$ operators of the `non-relativistic' theory.  
The exact analog for the current `relativistic' theory 
should be obvious, though there may still be particular 
interesting lessons to be learnt in a detailed analysis. 
Even though the explicit  new projective Hilbert space 
with the pseudo-unitary inner product is no mystery, 
a careful study of it from a physics point of view still 
has to be performed. Interestingly enough, it corresponds 
to a K\"ahler manifold of negative, instead of a positive, 
constant holomorphic sectional curvature (see Ref. \cite{082} 
and references therein). Seeing it as the noncommutative 
symplectic geometry, it is really a quantum model of the 
spacetime, instead of just of the phase space. The single 
particle phase space, an irreducible representation of the 
relativity symmetry, cannot be split into independent 
configuration and momentum spaces as in the classical limit, 
for $H_{\!\ssc R}(1,3)$ symmetry analyzed here, as well as  
for $H_{\!\ssc R}(3)$ group. This nonseparability is exactly 
analogous to that of the Minkowski spacetime representation 
of Lorentz symmetry (or any symmetry with Lorentz subgroup),  
whose space and time parts together form an irreducible 
representation, splitting into independent parts only in 
the Newtonian limit. The configuration space for
a single free particle in a theory of particle dynamics
is the only sensible physical notion of the model of the 
space behind it, as the space can only be understood as the 
collective of all possible positions a particle can occupy  
and be observed at. When that space is only a part of an 
integral whole, namely of an irreducible representation
of the fundamental relativity symmetry of the theory, it 
is full representation space that has to be considered, 
whether we call it is the spacetime or the phase space. 
So, quantum spacetime is the phase space, and we have
a solid model of a quantum spacetime with little
speculative element. It is just a Lorentz covariant
version of the quantum physical space model, with Minkowski 
four-vector position observables, generalizing the classical 
Minkowski spacetime with $x^\mu$ as coordinate observables
together with the momentum observable counterparts
binded as an irreducible object at the quantum level.
Hence our title.  

\bigskip\bigskip\noindent
\textbf{Appendix : The coherent states as seen from the Fock state basis }\\

The full set of Fock states as eigenstates for the covariant 
harmonic oscillator with the explicit wavefunction representation 
free from any divergence issue is given in \cite{083} under the 
same pseudo-unitary representation of $H_{\!\ssc R}(1,3)$, 
exactly with the Hermitian $\hat{X}_a$ and $\hat{P}_a$ and
$\hat{X}_{\!\ssc 0} = i \hat{X}_{\!\ssc 4}$ and
$\hat{P}_{\!\ssc 0} = i \hat{P}_{\!\ssc 4}$  anti-Hermitian.
The wavefunction given there are in the $x$-representation,
 functions of $x^a$ on which $\hat{X}_a$ and $\hat{P}_a$ 
act as operators $x_a$ and $-i\hbar \partial_{x^a}$.
In the following, we present the results under the 
convention and units used here. 

The ladder operators are defined as
\bea
\hat{a}_a=\hat{X}^{}_a+i\hat{P}_a\;,
\qquad  \hat{a}^\dag_a=\hat{X}_a-i\hat{P}_a\;;
\qquad\left[\hat{a}_a,\hat{a}^\dag_b \right]= 4\delta_{a b}\;,
\eea
The Fock states are simultaneous eigenstates
of $\hat{N}_a = \frac{1}{4}\hat{a}^\dag_a \hat{a}^{}_a$
and their sum $\hat{N}=\frac{1}{4}\hat{a}^\dag_a \hat{a}^a$,
\begin{equation} 
   \left|n\rra \equiv
    \left| n_{\! \ssc 1},n_{\! \ssc 2},n_{\! \ssc 3};n_{\! \ssc 4}\right\rangle  
=   \frac{1}{2^{n} \sqrt{\nfac}} \left(\hat{a}_{\!\ssc 1}^\dag \right)^{n_1} 
     \left(\hat{a}_{\!\ssc 2}^\dag\right)^{n_2}\left(\hat{a}_{\!\ssc 3}^\dag \right)^{n_3}
    \left(\hat{a}_{\!\ssc 4}^\dag \right)^{n_{4}}|0\rangle \;.
\end{equation}
In terms of the $\left|p^a,x^a \rra$ coherent states, for which 
$\hat{a}_b \left|p^a,x^a \rra = 2(x_b+ip_b) \left|p^a,x^a \rra$, 
we have the wavefunctions
\bea 
   \phi_n(p^a,x^a) \equiv   \lla p^a,x^a| n\rra
&=& \frac{1}{\sqrt{n_{\!\ssc 1}!\,n_{\!\ssc 2}!\,n_{\!\ssc 3}!\,n_{\!\ssc 4}!}}
   \left(x^{\ssc 1}-ip^{\ssc 1}\right)^{n_1} 
   \left(x^{\ssc 2}-ip^{\ssc 2} \right)^{n_2} 
\sea \qquad \qquad
  \times \left(x^{\ssc 3}-ip^{\ssc 3} \right)^{n_3} 
  \left(x^{\ssc 4}-ip^{\ssc 4} \right)^{n_4}\,
    e^{-\frac{x^a x_a+p^a p_a}{2}}\;.
\eea
Of course, they can also be obtained as
solutions to the eigenvalue equation of the $\hat{N}$
operator represented by the differential operator
based on the representation of $\hat{X}_a^{\!\ssc L}$ 
and $\hat{P}_a^{\!\ssc L}$. 

The Lorentz covariant picture of the results can be seen
with the Lorentz invariant inner product $\lla\!\lla\cdot|\cdot\rra\!\rra$, 
introduced as a set of basis bras, or functionals
$\lla\!\lla n \right|\right. \equiv (-1)^{n_{\! \ssc 4}} \!\lla n \right|$, 
on the basis states $\left.\left|n\rra\!\rra \equiv \left|n\rra$.
With the previously introduced 
$\left| n_{\! \ssc 0};n_{\! \ssc 1},n_{\! \ssc 2},n_{\! \ssc 3} \rra
 \equiv i^{n_{\! \ssc 4}} \left| n_{\! \ssc 1},n_{\! \ssc 2},n_{\! \ssc 3};n_{\! \ssc 4}\rra$ and $n_{\! \ssc 0} \equiv n_{\! \ssc 4}$,
and the language of pseudo-Hermitian quantum
mechanics sketched in Sec. \ref{sec3},  we have
\bea &&
\hat{a}^{\ssc 0}    \left.\!\left| n_{\! \ssc 0};n_{\! \ssc 1},n_{\! \ssc 2},n_{\! \ssc 3} \rra\!\rra
   = -i \hat{a}_{\!\ssc 4} \, i^{n_{\! \ssc 4}} \left.\!\left| n_{\! \ssc 1},n_{\! \ssc 2},n_{\! \ssc 3};n_{\! \ssc 4} \rra\!\rra
  =  2 \sqrt{n_{\! \ssc 0}}  \left. \!\left| n_{\! \ssc 0}-1;n_{\! \ssc 1},n_{\! \ssc 2},n_{\! \ssc 3} \rra\!\rra \;,
\sea
\hat{a}_{\ssc 0}^{\dag\!\!\dag} \left.\!\left| n_{\! \ssc 0};n_{\! \ssc 1},n_{\! \ssc 2},n_{\! \ssc 3} \rra\!\rra
   = i \hat{a}_{\!\ssc 4}^\dag \, i^{n_{\! \ssc 4}}  \left.\!\left| n_{\! \ssc 1},n_{\! \ssc 2},n_{\! \ssc 3};n_{\! \ssc 4} \rra\!\rra
  =  2 \sqrt{(n_{\! \ssc 0}+1)}   \left.\!\left| n_{\! \ssc 0}+1;n_{\! \ssc 1},n_{\! \ssc 2},n_{\! \ssc 3} \rra\!\rra  \;,
\eea
with $\frac{1}{4}\hat{a}_{\ssc 0}^{\dag\!\!\dag}  \hat{a}^{\ssc 0}
   = \hat{N}_{\!\ssc 0} =\hat{N}_{\!\ssc 4}$ and
  $\hat{N}=\frac{1}{4}\hat{a}_\mu^{\dag\!\!\dag}  \hat{a}^\mu$.
 $\hat{a}_\mu^{\dag\!\!\dag}$ denotes the 
${\mathcal{P}}_{\!\ssc 4}$-Hermitian conjugate of $\hat{a}_\mu$;
\bea
\hat{a}_\mu=\hat{X}_\mu+i\hat{P}_\mu\;,
\qquad  \hat{a}^{\dag\!\!\dag}_\mu=\hat{X}_\mu-i\hat{P}_\mu\;;
\qquad\left[\hat{a}_\mu,\hat{a}^{\dag\!\!\dag} _\nu \right]=4\eta_{\mu \nu}\;,
\eea
($\hat{a}^{\dag\!\!\dag}_i$ is identical with 
$\hat{a}^{\dag}_i$), which gives the expected form for 
a naive formulation of the Lorentz covariant problem 
since ${\mathcal{P}}_{\!\ssc 4}$-Hermitian conjugation 
is the Hermitian conjugation with respect to the inner 
product  $\lla\!\lla\cdot|\cdot\rra\!\rra$, and 
$\hat{X}_\mu$ and $\hat{P}_\mu$ are all
${\mathcal{P}}_{\!\ssc 4}$-Hermitian. However, the 
representation is really the pseudo-unitary in nature 
since the inner product is not positive definite. The 
$\left.\left|n\rra\!\rra$ states of odd $n_{\! \ssc 4}$ 
have timelike norm of $-1$; the ${\mathcal{P}}_{\!\ssc 4}$ 
operator was introduced as defined by  
${\mathcal{P}}_{\!\ssc 4} \left|n\rra = (-1)^{n_{\!4}} \left|n\rra$
giving $\lla\!\lla m|n\rra\!\rra = (-1)^{n_{\!4}} \delta_{mn}$

With the ${\mathcal{P}}_{\!\ssc 4}$-Hermitian (or 
${\mathcal{P}}_{\!\ssc 4}$-unitary formulation, one 
can see that the results look exactly like a usual 
unitary picture with the Fock states of the form
\begin{equation} 
    \left.\!\left| n_{\! \ssc 0};n_{\! \ssc 1},n_{\! \ssc 2},n_{\! \ssc 3} \rra\!\rra
=   \frac{1}{2^{n} \sqrt{n_{\! \ssc 0}!\, n_{\! \ssc 1}!\, n_{\! \ssc 2}! \, n_{\! \ssc 3}! }} 
   \left(\hat{a}_{\!\ssc 0}^{\dag\!\!\dag} \right)^{n_0}
    \left(\hat{a}_{\!\ssc 1}^{\dag\!\!\dag} \right)^{n_1} 
     \left(\hat{a}_{\!\ssc 2}^{\dag\!\!\dag}\right)^{n_2}
    \left(\hat{a}_{\!\ssc 3}^{\dag\!\!\dag} \right)^{n_3}
    |0\rangle\!\rangle \;.
\end{equation}
The true $H_{\!\ssc R}(1,3)$
coherent states can be introduced as
\bea
\left.\!\left|p^\mu,x^\mu\rra\!\rra
 &=& e^{-\frac{x^\mu x_\mu+p^\mu p_\mu}{2}}
   \sum\frac{1}{\sqrt{n_{\! \ssc 0}!\, n_{\! \ssc 1}!\, n_{\! \ssc 2}! \, n_{\! \ssc 3}! }}
   \left(x^{\ssc 0}+ip^{\ssc 0} \right)^{n_0}
   \left(x^{\ssc 1}+ip^{\ssc 1}\right)^{n_1} 
\sea \qquad \qquad
   \times \left(x^{\ssc 2}+ip^{\ssc 2} \right)^{n_2} 
   \left(x^{\ssc 3}+ip^{\ssc 3} \right)^{n_3} \,
   \left.\!\left| n_{\! \ssc 0};n_{\! \ssc 1},n_{\! \ssc 2},n_{\! \ssc 3} \rra\!\rra \;,
\eea
satisfying $\hat{a}^\nu \left.\!\left|p^\mu,x^\mu\rra\!\rra 
= 2 \left(x^{\nu}+ip^{\nu} \right) \left.\!\left|p^\mu,x^\mu\rra\!\rra$
and $\lla\!\lla p^\mu,x^\mu |p^\mu,x^\mu\rra\!\rra =1$.
Moreover, that is exactly the state given by
$V\!(p^\mu,x^\mu) = e^{i(p^\mu \hat{X}_\mu - x^\mu \hat{P}_\mu )}$ 
acting on $\left.\!\left|0\rra\!\rra$ 
as defined in the main text, or equivalently
\[
V\!(p^\mu,x^\mu) \left.\!\left|0\rra\!\rra
  = e^{-\frac{x^\mu x_\mu+p^\mu p_\mu}{2}} 
      e^{ \frac{  (x^{\nu}+ip^{\nu}) \hat{a}_{\nu}^{\dag\!\!\dag} }{2} }
    |0\rangle\!\rangle \;.
\]
We  have been doing nothing more than repeating
some standard calculations for a unitary picture,
most parts of which are not sensitive to the indefinite
nature of the norm. We have, however, 
\bea 
      \lla\!\lla p^\mu,x^\mu| n_{\! \ssc 0};n_{\! \ssc 1},n_{\! \ssc 2},n_{\! \ssc 3} \rra\!\rra
&=& \frac{1}{\sqrt{n_{\!\ssc 0}!\, n_{\!\ssc 1}!\,n_{\!\ssc 2}!\,n_{\!\ssc 3}!}}
    \left(x_{\ssc 0}-ip_{\ssc 0} \right)^{n_0}
   \left(x_{\ssc 1}-ip_{\ssc 1}\right)^{n_1} 
\sea \qquad \qquad \times 
   \left(x_{\ssc 2}-ip_{\ssc 2} \right)^{n_2} 
  \left(x_{\ssc 3}-ip_{\ssc 3} \right)^{n_3} \,
    e^{-\frac{x^\mu x_\mu+p^\mu p_\mu}{2}}\;,
\eea
which is really only about the distinction between an 
upper and a lower $0$ index to be traced carefully.
Taking that as the wavefunctions $\tilde\phi_n(p^\mu,x^\mu)$,
they differ from those one would obtain from a naive
unitary formulation exactly by having the factors of
$(x^\mu -i p^\mu)$ written as $(x_\mu -i p_\mu)$, which
really is about a factor of $(-1)^{n_{\! \ssc 0}}$.
However, the physical inner product between
two such wavefunctions  is very nontrivial.

It is important to note that $\left.\!\left|p^\mu,x^\mu\rra\!\rra$ 
states are different from the $\left.\!\left|p^a,x^a\rra\!\rra$
states. Recall that the latter are not normalized.
But each of them is also an eigenstate of $\hat{a}^{\ssc 0}$  
with the eigenvalue  $2(p^{\ssc 4}-i x^{\ssc 4})$.
Comparing the normalized $\left.\!\left|p^a,x^a\rra\!\rra$
with $\left.\!\left|p^\mu,x^\mu\rra\!\rra$ 
gives the consistent identification of the two for
$x^{\ssc 4}=-p^{\ssc 0}$ and $p^{\ssc 4}= x^{\ssc 0}$.
 Putting these relations into the standard resolution of
identity for the $\left|p^a,x^a\rra$ states gives the
 nontrivial inner product between wavefunctions 
given within the main text. 

\vspace*{.2in}
\noindent{\bf Acknowledgements \ }
S.B. thanks the Center for High Energy and High Field Physics,
National Central University for hospitality while the article
is being round up.
 O.K. and H.K.T. are partially supported by research grant 
number 107-2119-M-008-011
of the MOST of Taiwan.

\end{document}